\documentclass[a4paper,11pt]{article}

\usepackage{amssymb,amsmath,amsfonts}
\usepackage{graphicx}
\usepackage{color}
\usepackage{bm}
\usepackage{listings}
\usepackage{float}
\usepackage{todonotes}

\newcount\colveccount
\newcommand*\colvec[1]{
        \global\colveccount#1
        \begin{pmatrix}
        \colvecnext
}
\def\colvecnext#1{
        #1
        \global\advance\colveccount-1
        \ifnum\colveccount>0
                \\
                \expandafter\colvecnext
        \else
                \end{pmatrix}
        \fi
}

\newcommand{\x}{\mathbf{x}}
\newcommand{\y}{\mathbf{y}}

\newcommand{\ve}{\varepsilon}

\newcommand{\z}{\mathbf{z}}
\newcommand{\F}{\mathbf{F}}
\newcommand{\G}{\mathbf{G}}
\newcommand{\R}{\mathbb{R}}

\newlength{\figwidth}
\figwidth=\linewidth

\newcommand{\qed}{\nobreak \ifvmode \relax \else
      \ifdim\lastskip<1.5em \hskip-\lastskip
      \hskip1.5em plus0em minus0.5em \fi \nobreak
      \vrule height0.75em width0.5em depth0.25em\fi}

\usepackage{url}  
\usepackage{ifthen}  
\usepackage[utf8]{inputenc} 

\newcommand{\edita}[1]{{\color{black}{#1}}}
\newcommand{\editb}[1]{{\color{black}{#1}}}
\newcommand{\editc}[1]{{\color{black}{#1}}}
\newcommand{\editd}[1]{{\color{black}{#1}}}
\newcommand{\yp}[1]{{\color{black}{#1}}}
\newcommand{\ypb}[1]{{\color{black}{#1}}}

\begin{document}

\title{A multiple timescales approach to bridging spiking- and population-level dynamics}
\author{Youngmin Park\footnote{Corresponding author. Email yop6@pitt.edu} \,\,\& G. Bard Ermentrout \\ Department of Mathematics\\University of Pittsburgh\\ Pittsburgh PA 15260}

\maketitle

\begin{abstract}
A rigorous bridge between spiking-level and macroscopic quantities is an on-going and well-developed story for asynchronously firing neurons, but focus has shifted to include neural populations exhibiting varying synchronous dynamics. Recent literature has used the Ott--Antonsen ansatz (2008) to great effect, allowing a rigorous derivation of an order parameter for large oscillator populations. The ansatz has been successfully applied using several models including networks of Kuramoto oscillators, theta models, and integrate-and-fire neurons, along with many types of network topologies. In the present study, we take a converse approach: given the mean field dynamics of slow synapses, predict the synchronization properties of finite neural populations. The slow synapse assumption is amenable to averaging theory and the method of multiple timescales. Our proposed theory applies to two heterogeneous populations of N excitatory n-dimensional and N inhibitory m-dimensional oscillators with homogeneous synaptic weights. We then demonstrate our theory using two examples. In the first example we take a network of excitatory and inhibitory theta neurons and consider the case with and without heterogeneous inputs. In the second example we use Traub models with calcium for the excitatory neurons and Wang-Buzs{\'a}ki models for the inhibitory neurons. We accurately predict phase drift and phase locking in each example even when the slow synapses exhibit non-trivial mean-field dynamics.
\end{abstract}

\bigskip

\section{Lead Paragraph}
\editd{Mean field theory is one of the primary tools used by physicists and mathematicians working in neuroscience. For nearly five decades, the phenomenological derivation of neural field equations that describe population-level quantities such as the mean firing rate was sufficient, in part due to its success in reproducing the diverse spatio-temporal neural activity patterns of the neocortex. However, within the past two decades, theoretical studies have re-examined the derivation of the mean field models by starting at the microscopic, single-neuron spiking level. The spiking-level approach splits in two at the very beginning. In the first approach, the degree of synchrony of the population does not matter, and spiking neurons are assumed to be entirely asynchronous. This case is most similar to the classic mean field description. In the second approach, the degree of synchrony in the population does matter, and correlations between neurons are taken into account. The present study is most akin to the second approach, and extends upon existing results by introducing methods to reduce arbitrarily complex spiking models into a set of phase values that can be used to predict population synchrony.}

\section{Introduction}
Neural mean field models are a useful framework for studying mesoscopic and macroscopic spatio-temporal activity in the cortex. Examples include mammalian path integration, head direction tracking, visual hallucination, working memory, spatial object location, and object orientation \cite{coombes_2005_biolCyb,coombes_bumps_2005,folias_bressloff_2005_prl,breakspear2017dynamic}.

\editb{Existing studies derive macroscopic quantities starting at the spiking level, but require particular assumptions including asynchronous firing \cite{laing2014derivation,laing_exact_2015,renart2007mean} and Poisson statistics \cite{amit1997model,amit1997dynamics}. These studies contain no information about synchronization at the spiking level, which could underpin the loss or gain of power in electroencephalogram (EEG) frequency bands \cite{coombes2016next}.}

\editb{Recent studies relax the asynchronous firing assumption with the goal of predicting population synchrony, and have successfully used low-dimensional spiking models like the Kuramoto model \cite{so2008synchronization}, theta model \cite{ek84,coombes2016next}, Alder units \cite{roulet_average_2016}, and quadratic integrate-and-fire models \cite{montbrio_macroscopic_2015} to this end. These models are amenable to the Ott-Antonsen ansatz \cite{ott_long_2009}, which results in a complementary order parameter in addition to the mean field variables, e.g., firing rate.

In contrast to these studies, the goal of the present study is to derive a metric of synchrony at the spiking level for general oscillators. In particular, we derive a set of phase equations for each oscillator as a function of the mean field. Moreover, the existence of multiple timescales allows our approach to be converse to existing studies, where we derive the phase equations starting with the the mean field dynamics\yp{, which for the present study are the same as the mean synaptic variables.}}

We begin with a finite network of $N$, $n$-dimensional excitatory spiking neurons and $N$, $m$-dimensional inhibitory spiking neurons connected by slow synapses \yp{(The slow synaptic assumption is one of several that allows us to apply the theory of averaging and exploit the difference in time scales to get exact mean field equations for {\em finite} networks of neurons \cite{chandra_modeling_2017,ermentrout2010mathematical,ek84}. While, generally, synapses are fast, there are certain classes of excitatory and inhibitory synapses that are slow  (notably, NMDA and GABA$_\text{B}$ \cite{dayan2003theoretical}, and experimentally observed but unidentified synapses \cite{kleinfeld1990circuits}):}
\begin{align}
 \frac{d\x_i}{dt} &= \F^x(\x_i,s^x,s^y) + \edita{\ve \G_i^x(\x_i)}\label{eq:main1a},\\
 \frac{d\y_i}{dt} &= \F^y(\y_i,s^x,s^y) + \edita{\ve \G_i^y(\y_i)}\label{eq:main2a},\\
 \mu^x\frac{ds^x}{dt} &= \ve\left[-s^x + \frac{1}{\yp{N}}\sum_{i=1}^{\yp{N}}\yp{\sum_j} \delta(t-t\yp{^x_{i,j}})\right]\label{eq:main1b},\\
 \mu^y\frac{ds^y}{dt} &= \ve\left[-s^y + \frac{1}{\yp{N}}\sum_{i=1}^{\yp{N}}\yp{\sum_j} \delta(t-t\yp{^y_{i,j}})\right]\label{eq:main2b},
\end{align}
where $i=1\ldots,N$, \yp{$t^x_{i,j}$ ($t^y_{i,j}$) is the time of the $j^\text{th}$ spike of neuron $i$ in population $\x$ ($\y$)}.  The function \edita{$\F^k$ is the vector field for neuron $k=x,y$, and the function $\G_i^k$ represents heterogeneity in oscillator $i$ of vector field $k$.} \yp{The parameters $\mu_x/\ve,$ and $\mu_y/\ve$ represent the time constants of the $\x$ and $\y$ populations. We allow the $O(1)$ parameters $\mu_x,\mu_y$ to be different so that we can flexible vary the excitatory and inhibitory synaptic time scales.}. For conductance-based models, we take the spike time be the upwards zero-crossing of the membrane potential. The same notation holds for the inhibitory population $\y$. \yp{As we show next, our derivation only requires knowledge about whether or not any neuron spikes in a given population, thus we need not be precise about indexing the spike times by neuron number.}

\edita{Equations \eqref{eq:main1b} and \eqref{eq:main2b} \yp{represent the slow synaptic dynamics}.} \yp{The synaptic variable} $s^x$ ($s^y$) resets instantaneously (\yp{with small magnitude}) as $s^x \mapsto \bar s^x + \ve/(N\mu^x)$ ($s^y \mapsto \bar s^y + \ve/(N\mu^y)$) whenever $\x_i$ ($\y_i$) spikes. The small $\ve$ assumption results in slow timescale decay of the \yp{synaptic variables} $s^x,s^y$. \yp{Essentially, $s^x,s^y$ are low-pass filtered versions of the firing rates of the $x,y$ populations.} With these notations defined, we turn to the assumptions.

\begin{itemize}
 \item The term $\ve$ is small, $0 < \ve \ll 1$. Thus the synapses increment instantaneously with order $\ve$, but decay slowly between spikes.
 \item There is a separation of timescales into a ``fast'' time $t$ and a ``slow'' time $\tau = \ve t$.
 \item The mean synaptic values, $\bar s^x$ and $\bar s^y$, are constant \ypb{on the fast timescale} and differ at most by a small amount $O(\ve)$. These mean synapses represent the mean field values of Equations \eqref{eq:main1a}--\eqref{eq:main2b}. They must be close in value since they are the mean firing rates (frequencies) of the populations $x,y$ and in order for non-trivial phase-locking to occur between populations, their mean frequencies must be close.
 \item In the decoupled case \edita{without heterogeneity} ($\ve = 0$), we assume that there exists a $T$-periodic limit cycle $\Phi^k(t,\tau)$ satisfying Equations \eqref{eq:main1a},\eqref{eq:main2a} for a range of values $\bar s^k  \in [s^-,s^+]$ where $k=x,y$.
 \item Generally, $\bar s^k > 0$, and there \yp{might not exist a limit cycle} when $\bar s^k = 0$.
 \item The vector field dimensions are arbitrary: $\F^x,\edita{\G_i^x}: \R^n \times\R\times\R \rightarrow \R^n$ and $\F^y,\edita{\G_i^y}:\R^m \times \R \times \R \rightarrow \R^m$, where $n,m \geq 1$. 
\end{itemize}
These assumptions appear in a similar form in \ypb{Rubin et al.~2013} \cite{rubinrubin}. The authors show that when the mean synaptic values are constant with slow synaptic decay, the slow and strong coupling problem becomes a fast and weak coupling problem, and thus amenable to the classic phase reduction. However, we are \edita{also} interested in the case where the synaptic variables $s^x,s^y$ are slowly varying. \edita{Thus, we allow} that
\begin{itemize}
 \item There \edita{can} exist small amplitude (order $\sigma$, small) \ypb{slow} oscillations in the mean slow variables $\bar s^x, \bar s^y$ for some parameter values $\mu^x,\mu^y$ satisfying the mean field description of Equations \eqref{eq:main1a}--\eqref{eq:main2b}
 \begin{equation}\label{eq:meanfield}
  \begin{split}
   \mu^x \frac{d\bar s^x}{dt} &= \ve[-\bar s^x + \edita{\omega}^x(\bar s^x,\bar s^y)],\\
\mu^y \frac{d\bar s^y}{dt} &= \ve[-\bar s^y + \edita{\omega}^y(\bar s^x,\bar s^y)],
  \end{split}
 \end{equation}
 where $\edita{\omega}^k$ is the frequency of population $k=x,y$.
\end{itemize}

The goal of this paper is to derive a system of equations describing the phase locking properties of the network (Equations \eqref{eq:main1a}--\eqref{eq:main2b}). \yp{The equations will complement Equation \eqref{eq:meanfield}, which is the mean field formulation of Equations \eqref{eq:main1a}--\eqref{eq:main2b}}. The primary contribution of the paper is \editb{to show} that the phase reduction is valid when the synapses are slowly varying with small amplitude.

\yp{Before we start, we briefly summarize the derivation to follow. The first step of our analysis is a perturbation of order $\ve$ off of the uncoupled solution (Equations \eqref{eq:main1a}--\eqref{eq:main2b} with $\ve = 0$). The order $\ve$ terms of the expansion capture the weak and slow interactions of Equations \eqref{eq:main1a}--\eqref{eq:main2b}. In particular, we include slow timescale phase shifts $\theta_i^x(\tau)$ and $\theta_i^y(\tau)$, and explicitly write \yp{each synaptic variable} as its fixed mean value plus possible non-stationary terms. The goal of the derivation below is to determine the dynamics of the variables $\theta_i^x(\tau)$ and $\theta_i^y(\tau)$ by exploiting the separation in timescales. We find as expected that the phase dynamics exhibit all-to-all coupling, but less intuitively that small deviations of the mean synaptic values $\bar s^x, \bar s^y$ away from the fixed point $(\yp{s^*}, \yp{s^*})$ contribute to large phase drifts. With this summary in mind, we turn to the detailed derivation.}

\section{Derivation of the Phase Model}
We begin the reduction to phase oscillators with the ansatz
\begin{equation}\label{eq:ansatz}
\begin{split}
  \x_i(t,\tau) &= \x_i(t + \theta_i\yp{^x}(\tau),\yp{s^*}) = \Phi^x(t + \theta_i^x(\tau),\yp{s^*}) + \varepsilon \xi_i^x(t + \theta_i^x(\tau),\yp{s^*}) + O(\varepsilon^2),\\
  \y_i(t,\tau) &= \y_i(t + \theta_i\yp{^y}(\tau),\yp{s^*}) = \Phi^y(t + \theta_i^y(\tau),\yp{s^*}) + \varepsilon \xi_i^y(t + \theta_i^y(\tau),\yp{s^*}) + O(\varepsilon^2),\\
 s^x(t,\tau) &= \bar s^x(\tau) + \frac{\varepsilon}{N\mu^x} \sum_i f\left(t+\theta^x_i(\tau)\right) + O(\ve^2),\\
 s^y(t,\tau) &= \bar s\edita{^y}(\tau) + \frac{\varepsilon}{N\mu^y} \sum_i f\left(t+\theta^y_i(\tau)\right) + O(\ve^2),
\end{split}
\end{equation}
where $\bar s^k(\tau)$ is the slowly varying mean synaptic value for $k=x,y$, $f$ represents the effects of fast timescale, small-magnitude spikes on the synaptic variable, \yp{and $\theta_i^x(\tau)$, $\theta_i^y(\tau)$ are the slow timescale phase shifts due to the slow synaptic interactions}. \edita{Technically, \yp{the solutions}, $\x_i$, $\y_i$, \yp{uncoupled solutions}, $\yp{\Phi^k}$, and \yp{first order expansion term} $\xi_i^x$ have inputs of the form $(t+\theta_i(\tau),\bar s^x(\tau),\bar s^y(\tau))$, but because we evaluate these functions at the fixed mean value $\ypb{s^*}$, we abbreviate the notation of the redundant inputs by writing $(t+\theta_i(\tau),\yp{s^*})$}.

Using the periodicity of $s^k(t,\tau)$ on the fast timescale and small delta function impulses of order $\ve$, \edita{one can derive} $f$ explicitly as
\begin{align*}
 f(t+\theta^k(\tau)) &= \left [ \left(1 - (t+\theta^k(\tau))/T  \yp{\pmod{1}} \right) -1/2\right].
\end{align*}
We detail the \editc{calculations} in Appendix \ref{a:spiking}. \edita{For notational convenience, we do not write $f$ explicitly for the remainder of the derivation.}

Next, because the slow oscillations are small amplitude (order $\sigma$), we include an additional term in the expansion:
\begin{equation*}
 \begin{split}
   s^k(t,\tau) &= \bar s \edita{^k}(\tau) + \yp{s^* - s^*} + \edita{\frac{\varepsilon}{N\mu^\yp{k}}}\sum_j f\left(t+\theta^k_j(\tau)\right)\\
   &= \yp{s^*} + \sigma\left(\frac{\bar s^k(\tau) - \yp{s^*}}{\sigma}\right) +  \edita{\frac{\varepsilon}{N\mu^\yp{k}}} \sum_j f\left(t+\theta^k_j(\tau)\right),
 \end{split}
\end{equation*}
where $\yp{s^*}$ is the constant fixed mean values $\bar s^x = \bar s^y$. Plugging in Equation \eqref{eq:ansatz} into Equation \eqref{eq:main1a} and grouping in terms of small order ($\ve$ and $\sigma$) results in the system of equations,
\begin{align*}
 \frac{\partial \Phi^x}{\partial t} &= \F^x[\Phi^x(t + \theta_i^x(\tau)),\yp{s^*}],\\ 
 \frac{d\theta_i^x}{d\tau} &\frac{\partial \Phi^x}{\partial t}(t + \theta_i^x(\tau),\yp{s^*}) + \editc{\frac{d}{dt}\xi_i^{x}}(t + \theta_i^x(\tau),\yp{s^*})\\
 &=\F_{\Phi^x}^x(\Phi^x(t + \theta_i^x(\tau)),\yp{s^*})\xi_i^x(t + \theta_i^x(\tau),\yp{s^*})\\
 &\quad + \F^x_{s^x}(\Phi^x(t + \theta_i^x(\tau)),\yp{s^*})\left(\frac{\sigma}{\ve}\frac{\bar s^x(\tau) - \yp{s^*}}{\sigma}+ \edita{\frac{1}{N\mu^x}}\sum_j f\left(t+\theta^x_j(\tau)\right)\right)\\
 &\quad + \F^x_{s^y}(\Phi^x(t + \theta_i^x(\tau)),\yp{s^*})\left(\frac{\sigma}{\ve}\frac{\bar s^y(\tau) - \yp{s^*}}{\sigma}+ \edita{\frac{1}{N\mu^y}}\sum_j f\left(t+\theta^y_j(\tau)\right)\right)\\
 &\quad + \edita{\G_i^x(\Phi^x(t + \theta_i^x(\tau)))}.
\end{align*}
For an $n$-dimensional ($m$-dimensional) vector field $\F^x$ ($\F^y$), the derivative $\F^x_{\Phi^x}$ ($\F^y_{\Phi^y}$) represents the Jacobian matrix. Rewriting yields,
\begin{equation}\label{eq:eps_rhs}
\begin{split}
  &L\xi_i^x(t + \theta_i^x(\tau))\\
  &= \frac{d\theta_i^x}{d\tau} \frac{d\Phi^x}{dt}(t + \theta_i^x(\tau),\yp{s^*})\\
  &\quad -\F^x_{s^x}(\Phi^x(t + \theta_i^x(\tau)),\yp{s^*}) \left([\bar s^x(\tau) - \yp{s^*}]/\ve+\edita{\frac{1}{N\mu^x}}\sum_j f\left(t+\theta^x_j(\tau)\right)\right)\\
  &\quad -\F^x_{s^y}(\Phi^x(t + \theta_i^x(\tau)),\yp{s^*}) \left([\bar s^y(\tau) - \yp{s^*}]/\ve+\edita{\frac{1}{N\mu^x}}\sum_j f\left(t+\theta^y_j(\tau)\right))\right)\\
  &\quad - \edita{\G_i^x(\Phi^x(t + \theta_i^x(\tau)))},
\end{split}
\end{equation}
where
\begin{equation*}
 Lu \equiv -\editc{\frac{du}{dt}} + \F_{\Phi^x}^x(\Phi^x(t + \theta_i^x(\tau)),\yp{s^*})u.
\end{equation*}
Note that we have already collected terms in order $\ve$, so we have no need to keep the $\sigma/\sigma$ term and neglect it from now on. It is straightforward to show that the adjoint of \editd{$L$} is
\begin{align*}
\edita{ L^*v =v' + \left[\F_{\Phi^x}^x (\Phi^x(t + \theta_i^x(\tau)),\yp{s^*}) \right]^Tv.}
\end{align*}
We find that a function $\z^x$ in the nullspace of this adjoint operator satisfies
\begin{equation*}
\edita{ \frac{d\z^x}{dt}(t+\theta_i^x(\tau)) = - \left[\F_{\Phi^x}^x(\Phi^x(t + \theta_i^x(\tau)),\yp{s^*}) \right]^T \z^x(t+\theta_i^x(\tau)),}
\end{equation*}
and
\begin{equation*}
 \z^x \cdot \frac{d\Phi^x}{dt} = 1.
\end{equation*}
The function $\z^x$ is the same as the infinitesimal phase response curve of the oscillator $\Phi^x$ \cite{ermentrout2010mathematical}.

Next, we require the existence of a bounded periodic function $\xi_i^x$ satisfying Equation \eqref{eq:eps_rhs}. Because the operator $L$ has a closed range defined on the space of real-valued $T$-periodic functions, it follows that there exists a function $u$ satisfying $Lu=b$ if and only if $\langle u,v \rangle = 0$ for all $v$ in the nullspace of $L^*$ \cite{keener}, where
\begin{equation*}
 \langle u,v \rangle = \int_0^T u\cdot v\ dt.
\end{equation*}
Applying the existence condition directly to the right hand side of Equation \eqref{eq:eps_rhs} yields (with a bit of rearrangement)
\begin{align*}
 &\int_0^{T}\frac{d\theta_i^x}{d\tau} \frac{d\Phi^x}{dt}(t,\yp{s^*}) \cdot\z^x(t,\yp{s^*})\ dt\\
 = &\int_0^{T}\F^x_{s^x}(\Phi^x(t),\yp{s^*}) \cdot \z^x(t,\yp{s^*})\left( \frac{\bar s^x(\tau) - \yp{s^*}}{\ve}+\edita{\frac{1}{N\mu^x}}\sum_j f\left(t+\theta^x_j-\theta^x_i\right)\right)\ dt\\
 &+\int_0^{T}\F^x_{s^y}(\Phi^x(t),\yp{s^*}) \cdot \z^x(t,\yp{s^*})\left( \frac{\bar s^y(\tau) - \yp{s^*}}{\ve}+\edita{\frac{1}{N\mu^x}}\sum_j f\left(t+\theta^y_j-\theta^x_i\right)\right)\ dt\\
 &+\edita{\int_0^{T}\G_i^x(\Phi^x(t),\yp{s^*}) \cdot \z^x(t,\yp{s^*})\ dt}.
\end{align*}
Simplifying and rewriting, we arrive at the phase equations:
\begin{equation}\label{eq:theta_x}
 \begin{split}
   \frac{d\theta_i^x}{d\tau} &= [\bar s^x(\tau) - \yp{s^*}]\beta^{xx}/\ve + [\bar s^y(\tau) - \yp{s^*}]\beta^{xy}/\ve + \edita{B_i^x}\\
&\quad + \edita{\frac{1}{N\mu^x}}\sum_{j=1}^N H^{xx}(\theta_j^x(\tau) - \theta_i^x(\tau)) + \edita{\frac{1}{N\mu^x}}\sum_{j=1}^N H^{xy}(\theta_j^y(\tau) - \theta_i^x(\tau)),
 \end{split}
\end{equation}
where
\begin{align*}
\beta^{xy} &= \frac{1}{T}\int_0^{T}\F^x_{s^y}(\Phi^x(t),\yp{s^*})\cdot\z^x(t,\yp{s^*})dt,\\
\beta^{xx} &= \frac{1}{T}\int_0^{T}\F^x_{s^x}(\Phi^x(t),\yp{s^*})\cdot\z^x(t,\yp{s^*})dt,\\
H^{xx}(\phi) &= \frac{1}{T}\int_0^{T}\F^x_{s^x}(\Phi^x(t),\yp{s^*})\cdot\z^x(t,\yp{s^*})f(t+\phi)dt,\\
H^{xy}(\phi) &= \frac{1}{T}\int_0^{T}\F^x_{s^y}(\Phi^x(t),\yp{s^*})\cdot\z^x(t,\yp{s^*})f(t+\phi)dt,\\
\edita{B_i^x} &= \edita{\frac{1}{T}\int_0^{T}\G_i^x(\Phi^x(t),\yp{s^*}) \cdot \z^x(t,\yp{s^*})\ dt}.
\end{align*}
\editb{The vigilant reader may notice a possible issue with the term $(\bar s^x(\tau) - \yp{s^*})/\ve$, where $\ve$ is small. We require that the deviations of $\bar s^x(\tau)$ from $\yp{s^*}$ to be small, in particular to be of order $\sigma$. In our derivation, the order $\sigma$ term cancels so that we can treat the difference $(\bar s^x(\tau) - \yp{s^*})$ as order $\ve$. Thus, the term $(\bar s^x(\tau) - \yp{s^*})/\ve$ is order $O(1)$.}

Using the same arguments, we arrive at the phase equations for the $y$ population,
\begin{equation}\label{eq:theta_y}
\begin{split}
\frac{d\theta_i^y}{d\tau} &= [\bar s^x(\tau) - \yp{s^*}]\beta^{yx}/\ve + [\bar s^y(\tau) - \yp{s^*}]\beta^{yy}/\ve + \edita{B_i^y}\\
&\quad + \edita{\frac{1}{N\mu^y}}\sum_{j=1}^N H^{yx}(\theta_j^x(\tau) - \theta_i^y(\tau)) + \edita{\frac{1}{N\mu^y}}\sum_{j=1}^N H^{yy}(\theta_j^y(\tau) - \theta_i^y(\tau)),
\end{split}
\end{equation}
where
\begin{align*}
\beta^{yx} &= \frac{1}{T}\int_0^{T}\F^y_{s^x}(\Phi^y(t),\yp{s^*})\cdot\z^y(t,\yp{s^*})dt,\\
\beta^{yy} &= \frac{1}{T}\int_0^{T}\F^y_{s^y}(\Phi^y(t),\yp{s^*})\cdot\z^y(t,\yp{s^*})dt,\\
H^{yx}(\phi) &= \frac{1}{T}\int_0^{T}\F^y_{s^x}(\Phi^y(t),\yp{s^*})\cdot\z^y(t,\yp{s^*})f(t+\phi)dt,\\
H^{yy}(\phi) &= \frac{1}{T}\int_0^{T}\F^y_{s^y}(\Phi^y(t),\yp{s^*})\cdot\z^y(t,\yp{s^*})f(t+\phi)dt,\\
\edita{B_i^y} &= \edita{\frac{1}{T}\int_0^{T}\G_i^y(\Phi^y(t),\yp{s^*}) \cdot \z^y(t,\yp{s^*})\ dt}.
\end{align*}
Note that in the phase equations \eqref{eq:theta_x} and \eqref{eq:theta_y}, the synaptic variables are exogenous and do not depend on the microscopic solutions -- only the microscopic solutions depend on the mean field. Thus, the microscopic dynamics are fully described by properties of the individual oscillators (the iPRC $\z^k$, the vector field $\F^k$), and the mean synaptic variables $\bar s^k$.

When analyzing solutions, we use the phase differences $\phi_i^x = \theta_i^x - \theta_1^x$, $\phi_i^y = \theta_i^y - \theta_1^y$, where $i=1,\ldots,N$, and $\phi^z = \theta_1^y - \theta_1^x$. By definition, $\phi_1^x = \phi_1^y = 0$ and $d\phi_1^x/d\tau = d\phi_1^y/d\tau = 0$, so we only plot phase differences for $j > 1$. \edita{As we have shown in our derivation, our theory tolerates order $\ve$ heterogeneities in the vector fields}. The phase difference dynamics are then
\begin{equation}\label{eq:phase_diff_x}
\begin{split}
\edita{N\mu^x}\frac{d\phi_i^x}{d\tau} &=  \sum_{j=1}^N \left[H^{xx}\left(\phi_j^x - \phi_i^x\right) - H^{xx}\left(\phi_j^x\right)\right]+\editb{B^x_i - B^x_1}\\
&\quad + \sum_{j=1}^N \left[H^{xy}\left(\phi_j^y - \phi_i^x+\phi^z\right) - H^{xy}\left(\phi_j^y+\phi^z\right) \right],
\end{split}
\end{equation}
\begin{equation}\label{eq:phase_diff_y}
 \begin{split}
\edita{N\mu^y}\frac{d\phi_i^y}{d\tau} &=  \sum_{j=1}^N\left[ H^{yy}\left(\phi_j^y - \phi_i^y\right) - H^{yy}\left(\phi_j^y\right) \right]+\editb{B^y_i-B^y_1}\\
&\quad +\sum_{j=1}^N \left[H^{yx}\left(\phi_j^x - \phi_i^y-\phi^z\right) - H^{yx}\left(\phi_j^x-\phi^z\right)\right],
 \end{split}
\end{equation}
\begin{equation}\label{eq:phase_diff_z}
 \begin{split}
\frac{d\phi^z}{d\tau} &=  [\bar s^x(\tau) - \yp{s^*}](\beta^{yx}-\beta^{xx})/\ve + [\bar s^y(\tau) - \yp{s^*}](\beta^{yy} - \beta^{xy})/\ve\\
& \quad + \edita{\frac{1}{N\mu^y}}\sum_{j=1}^N\left[ H^{yx}\left(\phi_j^x -\phi^z\right) + H^{yy}\left(\phi_j^y\right)\right]\\
& \quad - \edita{\frac{1}{N\mu^x}}\sum_{j=1}^N\left[ H^{xx}\left(\phi_j^x\right) + H^{xy}\left(\phi_j^y+\phi^z\right) \right]\\
& \quad + \editb{B^y_1 - B^x_1}
 \end{split}
\end{equation}
where $i=1,\ldots,N$. When the mean synaptic variables are slowly varying, the terms $\bar s^k(\tau) - \yp{s^*}$ in the right hand side of $d\phi^z/d\tau$ are what contribute to large phase drifts between the populations.

To aid in the numerics and analysis, we make note of some facts, starting with the relationship between constant mean synapses and frequency.

\subsection{Relationship Between Constant Mean Synapses and Frequency}
Suppose that the mean $\bar s^k$ is constant and \yp{$\bar s^k = s^*$}. Recall that \yp{for the synaptic variable $s^k$} following a spike,
\begin{equation*}
 s^k(t) = s^k(0) e^{-\ve t/\mu^k}, t < T^-,
\end{equation*}
where $T^-$ is the period of the fast oscillator up to and not including the spike. We may determine the initial condition by solving
\begin{equation*}
 s^k(T^+) = s^k(0) e^{-\ve T/\mu^k} + \ve/\mu^k =  s^k(0),
\end{equation*}
which yields
\begin{equation*}
 s^k(0) = \frac{\ve}{\mu^k} \frac{1}{1-e^{-\ve T/\mu^k}}.
\end{equation*}
Taking the mean value of $s^k(t)$ over one period,
\begin{equation*}
 \bar s^k = \frac{1}{T} \int_0^T s^k(t) dt,
\end{equation*}
we find that
\begin{equation}\label{eq:mean=s}
 \bar s^k = \frac{1}{T}.
\end{equation}
That is, $\yp{s^* = \bar s^k}$ is the same as the fast frequency.

\edita{\subsection{Fourier Approximation}
Because the domain of each function $H^{jk}$ is periodic, we can use a Fourier series approximation to make the numerics tractable. We extract the Fourier coefficients using the fast Fourier transform (FFT) and construct an approximation by writing
\begin{equation*}
 H^{jk}(x) = \sum_{n=0}^\editb{M} (a_n \cos(n x/T) + b_n \sin(n x/T)).
\end{equation*}
All right hand sides can be written as a sum of sines and cosines, thus amenable to a bifurcation analysis using \texttt{XPPAUTO}. Constructing the Jacobian matrix using derivatives of $H^{jk}$ is also straightforward, since we only need to take the derivative of sines and cosines:
\begin{equation*}
 \frac{d H^{jk}}{dx}(x) = \sum_{n=0}^\editb{M} \left[-na_n\cos(n x/T)/T + n b_n \sin(n x/T)/T\right].
\end{equation*}
See Tables \ref{tab:fourier_theta}, \ref{tab:fourier_tbwb} for the values of the Fourier coefficients.
}
\section{Results}

We now turn to the simulation of neural models to test our theory. We begin by considering a population of excitatory and inhibitory theta neurons \cite{ek84} and look at two cases: first when the mean synaptic values are fixed, and second when the mean synaptic values are slowly varying with small amplitude about a fixed point. In the first case we show the existence and stability of various phase locked solutions. In the second case we use numerics to demonstrate the accuracy of our phase model.

We conclude by repeating the same comparison using biophysically realistic models. The models we consider are excitatory Traub models with calcium \cite{traub1982simulation}, and inhibitory Wang-Buzs{\'a}ki models \cite{wang1996gamma}.

\subsection{Theta Neurons}
Consider a network of excitatory and inhibitory theta neurons with all-to-all coupling,
\begin{equation}\label{eq:theta_network}
\begin{split}
 \frac{dx_j}{dt} &= \pi(1-\cos(x_j) + (1+\cos(x_j))[a^x+b^x s^x - c^x s^y]),\\
 \frac{dy_j}{dt} &= \pi(1-\cos(y_j) + (1+\cos(y_j))[a^y+b^y s^x - c^y s^y]),\\
 \mu^x \frac{ds^x}{dt} &= \ve \left[-s^x + \frac{1}{N} \sum_j \delta(x_j - \pi)\right],\\
 \mu^y \frac{ds^y}{dt} &= \ve \left[-s^y + \frac{1}{N} \sum_j \delta(y_j - \pi)\right],
\end{split}
\end{equation}
\yp{where $a^{x,y}, b^{x,y}$, and $c^{x,y}$ are positive constants chosen such that the main assumptions of this paper are satisfied. In this system, the dynamics of both populations are virtually identical, but the distinguishing features are the parameters $b^x,b^y$ and $c^x,c^y$. Whenever a neuron in population $\x$ ($\y$) spikes, it will positively (negatively) contribute to the dynamics of both populations, due to the $b^{x},b^{y}>0$ ($-c^{x},-c^{y}<0$) terms.}
Given values $\bar s^x$ and $\bar s^y$, the period of the oscillators is given by the solution to the integral
\begin{equation*}
 T^k = \frac{1}{\pi}\int_{-\pi}^{\pi}\frac{d x}{1-\cos(x) + (1+\cos(x))I^k}, \quad k=x,y,
\end{equation*}
where $I^k = a^k+b^k \bar s^x - c^k \bar s^y$. The \editd{reciprocal} of the solution is the frequency,
\begin{equation*}
 f^k(\bar s^x,\bar s^y) = \sqrt{\editc{[I^k]_+}},
\end{equation*}
where \editc{$[x]_+ = \max\{0,x\}$}. Thus, the averaged dynamics are
\begin{align}
\mu^x \frac{ds^x}{dt} &= \ve\left(-s^x + \sqrt{\editc{[I^x]_+}}\right)\label{eq:theta_avg1},\\
\mu^y \frac{ds^y}{dt} &= \ve\left(-s^y + \sqrt{\editc{[I^y]_+}}\right)\label{eq:theta_avg2}.
\end{align}
For this system, the limit cycle and iPRC are, respectively,
\begin{align*}
 \Phi^i(t,\yp{s^*}) &\equiv \Phi(t,\yp{s^*}) = 2 \arctan(\yp{s^*} \tan(\yp{s^*}\pi (t+T/2)),\\
 Z^i(t,\yp{s^*}) &\equiv Z(t,\yp{s^*}) = [\cos^2(\yp{s^*}\pi (t+T/2)) + \yp{(s^*)}^2\sin^2(\yp{s^*}\pi (t+T/2))]/(2\yp{(s^*)}^2\pi),
\end{align*}
where $\yp{s^*}$ is the fixed point $\bar s^x = \bar s^y$. To compute the $H$ functions, we note that
\begin{align*}
 \F^x_{s^x}(\Phi(t,\yp{s^*}),\yp{s^*}) &= b^x\pi[1 + \cos(\Phi(t,\yp{s^*}))],\\
 \F^x_{s^y}(\Phi(t,\yp{s^*}),\yp{s^*}) &= -c^x\pi[1 + \cos(\Phi(t,\yp{s^*}))],\\
 \F^y_{s^x}(\Phi(t,\yp{s^*}),\yp{s^*}) &= b^y\pi[1 + \cos(\Phi(t,\yp{s^*}))],\\
 \F^y_{s^y}(\Phi(t,\yp{s^*}),\yp{s^*}) &= -c^y\pi[1 + \cos(\Phi(t,\yp{s^*}))].
\end{align*}
Thus the $H$ functions of Equations \eqref{eq:theta_x} and \eqref{eq:theta_y} for this system are given by
\begin{equation}\label{eq:h_fun_theta}
 \begin{split}
 H^{xx}(\phi) &= \frac{b^x\pi}{T\mu^x}\int_0^{T} Z(t,\yp{s^*})[1 + \cos(\Phi(t,\yp{s^*}))]f(t + \phi) dt,\\
 H^{xy}(\phi) &= -\frac{c^x\pi}{T\mu^y}\int_0^{T} Z(t,\yp{s^*})[1 + \cos(\Phi(t,\yp{s^*}))]f(t + \phi) dt,\\
 H^{yx}(\phi) &= \frac{b^y\pi}{T\mu^x}\int_0^{T} Z(t,\yp{s^*})[1 + \cos(\Phi(t,\yp{s^*}))]f(t + \phi) dt,\\
 H^{yy}(\phi) &= -\frac{c^y\pi}{T\mu^y}\int_0^{T} Z(t,\yp{s^*})[1 + \cos(\Phi(t,\yp{s^*}))]f(t + \phi) dt.\\
 \end{split}
\end{equation}
\begin{figure}[t]
\centering
 \includegraphics[width=.75\textwidth]{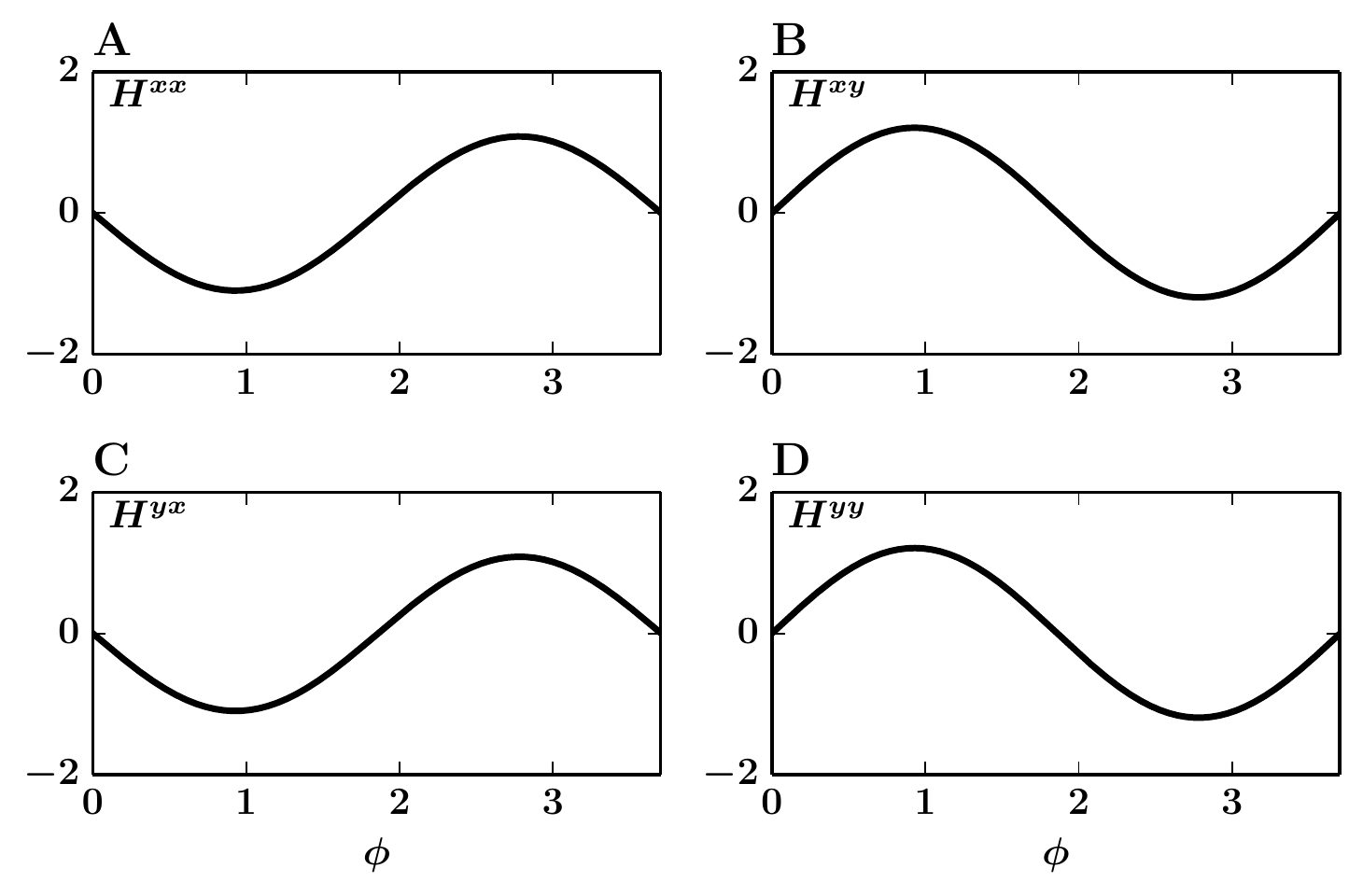}
 \caption{Example $H$-functions of the theta model. \edita{Panel A: $H^{xx}$. Panel B: $H^{xy}$. Panel C: $H^{yx}$. Panel D: $H^{yy}$.} In all panels, parameter values are $a^x=a^y=0.1$, $b^x=b^y=1$, $c^x=c^y=1.1$, and $\mu^x=\mu^y=1$.}\label{fig:h_theta}
\end{figure}
\edita{We show examples of the $H$-functions in Figure \ref{fig:h_theta}}. For clarity in the calculations to follow, we define a new function $\hat H^{jk}$ in order to write the parameters explicitly.
\begin{align*}
 b^x\hat H^{xx}(\phi)/\mu^x &=H^{xx}(\phi),\\
 -c^x\hat H^{xy}(\phi)/\mu^y &= H^{xy}(\phi),\\
 b^y\hat H^{yx}(\phi)/\mu^x &= H^{yx}(\phi),\\
 -c^y\hat H^{yy}(\phi)/\mu^y &= H^{yy}(\phi).
\end{align*}
Note that the slope of $\hat H^{ky}$ is the opposite of the slope of $H^{ky}$ for $k=x,y$.

\begin{figure}[ht!]
\centering
 \includegraphics[width=.75\textwidth]{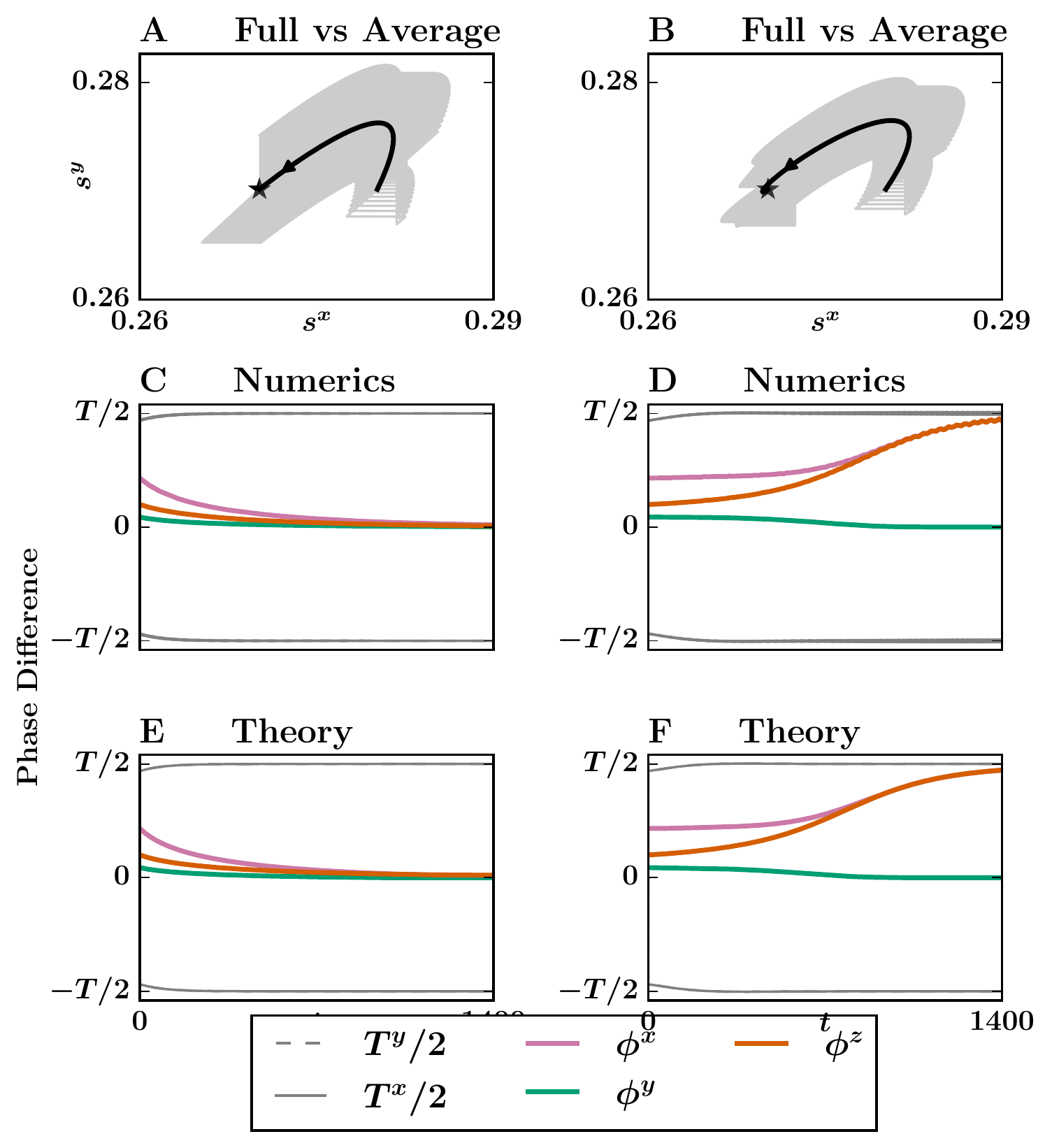}
 \caption{Mean field and microscopic behavior with constant mean synapses. \edita{Panel A}: mean field \yp{synaptic variables ($\bar s^{x,y}$}, black) plotted on top of the \yp{full network synaptic variables ($s^{x,y}$}, gray). The black star denotes a stable fixed point. \edita{Panel C}: phase difference at the spiking level in the full simulation. The estimated anti-phase value is shown in gray solid ($T^x/2$) and gray dashed ($T^y/2$) (they happen to overlap substantially and are almost impossible to distinguish). \edita{Panel E}: phase difference using our proposed phase reduction. Parameters are the same as in Figure \ref{fig:h_theta} with $\mu^x=\mu^y=1$ and \editd{$\ve= 0.01$}. In the right column \edita{(panels B, D, and F)}, we increase $\mu^y=1$ to $\mu^y = 1.4$ and plot data in the same order as \edita{panels A, C, and E, respectively}, .}\label{fig:micro_vs_macro}
\end{figure}

Using the tools developed up to this point, we can begin to explore the limitations of the mean field description and test if our phase reduction successfully captures the spiking-level synchronization. For a rudimentary demonstration of a mean field description that carries no information about microscopic dynamics, we direct our attention to Figure \ref{fig:micro_vs_macro}.

In this figure, we simulate a small network of $N=2$ excitatory and $N=2$ inhibitory theta neurons \edita{(for simplicity we define $\phi^x = \theta_2^x-\theta_1^x$, $\phi^y = \theta_2^y - \theta_1^y$, and $\phi^z = \theta_1^y - \theta_1^x$). In the left column, panel A represents the dynamics of the mean field description ($\bar s^{x,y}$) \editc{overlaid} on the full network synaptic variables ($s^{x,y}$) plotted in gray. Panel C shows the synchronization properties of the spiking model, and and panel E shows our proposed theory. The theory correctly predicts synchronization of all oscillators. In the right column, panels B, D, and F show the mean field model, spiking model, and proposed theory, respectively. All panels A--F use the same parameters as in Figure \ref{fig:h_theta}, except for the right column (panels B, D, and F) where we take $\mu^y=1.4$. The antiphase lines representing $T^x/2$ (gray solid) and $T^y/2$ (gray dashed) are hard to distinguish because they happen to \editd{nearly} coincide.}

Strikingly, we observe changes in the microscopic synchronization despite virtually no change in the mean field description. There is a slight quantitative change in the mean field descriptions when $\mu^y$ changes from $\mu^y=1$ to $\mu^y=1.4$. In particular, when $\mu^y=1$, the fixed point is an asymptotically stable node with real negative eigenvalues. When we increase $\mu^y$ to $\mu^y=1.4$, the fixed point remains stable but becomes a spiral node with small imaginary eigenvalues. Thus, the fixed point remains asymptotically stable and a numerical analysis of the mean field does not reveal any bifurcation points. Moreover, prior knowledge of this quantitative difference gives no indication with regards to the change in synchronization properties.

\subsection{Existence of Synchronous Solutions}
The existence of synchronous solutions \editc{is} straightforward to show. Generically, the synchronous solution $\phi^x_i=\phi^y_i=0$ for $i=2,\ldots,N$ exists independent of $\phi^z$ (all right-hand-side terms cancel with these values), even when the mean synaptic variables are slowly varying. However, in this slowly varying case, there is synchrony within each excitatory or inhibitory population, but not between populations -- the variable $\phi^z$ undergoes a large phase drift.

Similar solutions are just as straightforward to show. For example, $\phi^x_i = 0$ ($\phi^y_i=0$) for $i=1,\ldots,N$ gives us $d\phi_i^x/d\tau=0$ ($d\phi_i^y/d\tau=0$) independent of the dynamics of $\phi^y_i$ ($\phi^x_i$) and $\phi^z$. Thus, it is possible for the excitatory (inhibitory) population to remain synchronous despite a phase drift between populations and possibly asynchronous behavior in the inhibitory (excitatory) population. \editd{This behavior is not restricted} to the theta model and exist\yp{s} generically.

\subsection{Existence and Stability of Phase-Locked Solutions (Fixed Mean)}
We now determine the stability of a given phase-locked solution to Equations \eqref{eq:phase_diff_x},\eqref{eq:phase_diff_y}, and \eqref{eq:phase_diff_z} in the case of a fixed mean. To this end, we begin with the most general case of a generic phase-locked solution and construct the Jacobian matrix using the following derivatives:
\begin{equation*}
 \left[ \frac{\partial}{\partial \phi_2^x} \cdots \frac{\partial }{\partial \phi_N^x},\,\,\, \frac{\partial}{\partial \phi_2^y} \cdots \frac{\partial}{\partial \phi_N^y}, \,\,\,\frac{\partial}{\partial \phi^z} \right].
\end{equation*}
First, consider the partial derivatives with respect to $\phi_k^x$, $\phi_k^y$, and $\phi^z$ of the right hand side of $d\phi_i^x/d\tau$:
\begin{equation}\label{eq:jacx}
\begin{split}
N\frac{\partial}{\partial \phi_k^x}\frac{d\phi_i^x}{d\tau} &= \sum_{j=1}^N\left[H^{xx}_{\phi}(\phi_j^x - \phi_i^x) (\delta_{jk} - \delta_{ik}) - H^{xx}_{\phi}(\phi_j^x) \delta_{jk}\right],\\
&\quad+\sum_{j=1}^N H^{xy}_{\phi}(\phi_j^y-\phi_i^x+\phi^z)(-\delta_{ik}),\\
N\frac{\partial}{\partial \phi_k^y}\frac{d\phi_i^x}{d\tau} &= \sum_{j=1}^N\left[H^{xy}_{\phi}(\phi_j^y - \phi_i^x + \phi^z)\delta_{jk} - H^{xy}_{\phi}(\phi_j^y + \phi^z)\delta_{jk}\right],\\
N\frac{\partial}{\partial \phi^z}\frac{d\phi_i^x}{d\tau} &= \sum_{j=1}^N\left[H^{xy}_{\phi}(\phi_j^y - \phi_i^x + \phi^z) - H^{xy}_{\phi}(\phi_j^y + \phi^z)\right].
\end{split}
\end{equation}
The \edita{Kronecker} delta functions are defined as
\begin{equation*}
\delta_{ij} = \begin{cases}
  1, & \text{if } i = j, \\
  0, & \text{else},
\end{cases}
\end{equation*}
and $H_\phi$ denotes the derivative of $H$ with respect to its independent variable. Next, the partials with respect to $\phi_k^x$, $\phi_k^y$, $\phi^z$ of the right hand side of $d\phi_i^y/d\tau$:
\begin{equation}\label{eq:jacy}
\begin{split}
N\frac{\partial}{\partial \phi_k^x}\frac{d\phi_i^y}{d\tau} &= \sum_{j=1}^N\left[H^{yx}_{\phi}(\phi_j^x - \phi_i^y - \phi^z)\delta_{jk} - H^{yx}_{\phi}(\phi_j^x - \phi^z)\delta_{jk}\right],\\
N\frac{\partial}{\partial \phi_k^y}\frac{d\phi_i^y}{d\tau} &= \sum_{j=1}^N H^{yx}_{\phi}(\phi_j^x - \phi_i^y - \phi^z)(-\delta_{ik}),\\
 &\quad+\sum_{j=1}^N\left[H^{yy}_{\phi}(\phi_j^y - \phi_i^y)(\delta_{jk} - \delta_{ik}) - H^{yy}_{\phi}(\phi_j^y)\delta_{jk}\right],\\
N\frac{\partial}{\partial \phi_k^z}\frac{d\phi^y}{d\tau} &= \sum_{j=1}^N\left[H^{yx}_{\phi}(\phi_j^x - \phi_i^y - \phi^z)(-1) - H^{yx}_{\phi}(\phi_j^x - \phi^z)(-1)\right].
\end{split}
\end{equation}
Finally, the partials with respect to $\phi_k^x$, $\phi_k^y$, $\phi^z$ of the right hand side of $d\phi^z/d\tau$:
\begin{equation}\label{eq:jacz}
\begin{split}
N\frac{\partial}{\partial \phi_k^x}\frac{d\phi^z}{d\tau}&=\sum_{j=1}^N\left[H^{yx}_{\phi}(\phi_j^x - \phi^z)\delta_{jk} -  H^{xx}_{\phi}(\phi_j^x)\delta_{jk}\right],\\
N\frac{\partial}{\partial \phi_k^y}\frac{d\phi^z}{d\tau}&=\sum_{j=1}^N\left[H^{yy}_\phi(\phi_j^y)\delta_{jk} - H^{xy}_\phi(\phi_j^y + \phi^z)\delta_{jk}\right],\\
N\frac{\partial}{\partial \phi_k^z}\frac{d\phi^z}{d\tau}&=\sum_{j=1}^N\left[ H^{yx}_{\phi}(\phi_j^x - \phi^z)(-1) - H^{xy}_{\phi}(\phi_j^x + \phi^z)\right].
\end{split}
\end{equation}
The synchronous solution, $\phi_i^y = \phi_i^x = 0$ is most straightforward to analyze. In this case, all off-diagonal terms cancel except the last row, so the Jacobian matrix is \editc{lower-}triangular with diagonal entries
%
\begin{equation}\label{eq:sync_eigenvalues}
 \begin{split}
 NJ_{ii} &= -b^x \hat H^{xx}_\phi(0)/\mu^x + c^x \hat H^{xy}_\phi(0)/\mu^y,\quad i=1,\ldots, N-1\\
 NJ_{ii} &= -b^y \hat H^{yx}_\phi(0)/\mu^x + c^y \hat H^{yy}_\phi(0)/\mu^y,\quad i=N,\ldots,2N-2\\
 NJ_{2N-1,2N-1} &= -b^y \hat H^{yx}_\phi(0)/\mu^x +c^x\hat H^{xy}_\phi(0)/\mu^y.
 \end{split}
\end{equation}

These entries form the eigenvalues of the Jacobian matrix. We have seen in Figure \ref{fig:h_theta} that $H^{kx}(0)$ has negative slope for $k=x,y$ (panels A,C) and $H^{ky}(0)$ has positive slope (and hence negative slope for $\hat H^{ky}(0)$) for $k=x,y$ (panels B,D). Then, for $\mu^y$ sufficiently large, the negative contributions from functions $H^{ky}$ are small and the eigenvalues may become positive, indicating a loss of stability to the synchronous solution. This loss of stability confirms our observation in Figure \ref{fig:micro_vs_macro}.

\begin{figure}
\centering
 \includegraphics[width=.4\textwidth]{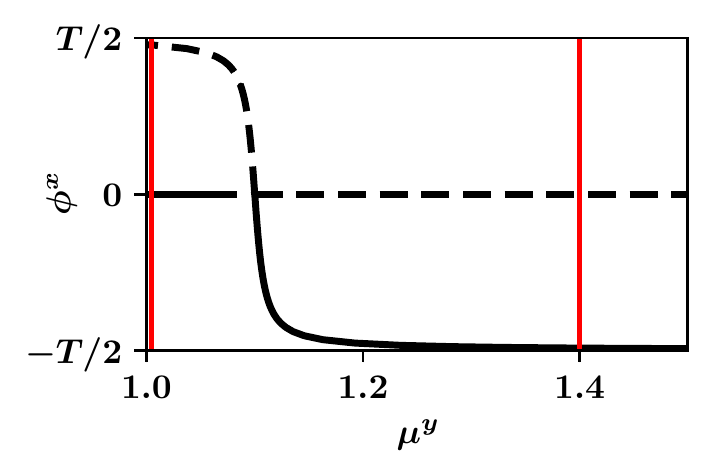}
 \caption{\editc{Stability of the fixed point taking the form $(\phi^x,0,0)$. The solution $\phi^x=0$ is stable when $\mu^y = 1$, and becomes unstable as $\mu^y$ increases through $\mu^y=1.1$. When $\mu^y=1.4$, the stable solution is of the form $(-T/2,0,0)$ indicating anti-phase solutions are stable in the excitatory population.}}\label{fig:one_nonsync_existence_stability}
\end{figure}

\editc{We found that non-synchronous fixed point attractors of this network take the form $(\phi^x,0,0)$, or $(\phi^x,0,\phi^z)$. For the remainder of this subsection, we analyze the existence and stability of fixed points starting with the synchronous solution $\phi^x=\phi^y=\phi^z=0$.

We can show that the bifurcation point occurs at $\mu^y=1.1$ by writing down the eigenvalues of this system (Equation \eqref{eq:sync_eigenvalues} with $N=2$):
\begin{align*}
 \lambda_1 &= \left[ -H^{xx}_\phi(0) - H^{xy}_\phi(0) \right],\\
 \lambda_2 &= \left[ -H^{yx}_\phi(0) - H^{yy}_\phi(0) \right],\\
 \lambda_3 &= \left[ -H^{yx}_\phi(0) - H^{xy}_\phi(0) \right].
\end{align*}

These $H$ functions are identical except for the choice of parameters $b^x=b^y=1$, and $c^x=c^y=1.1$ (Equation \eqref{eq:h_fun_theta}). By inspection, the eigenvalues are zero when when $\mu^x = 1$ and $\mu^y=1.1$ indicating a change of stability at $\mu^y=1.1$. This change in stability is shown in Figure \ref{fig:one_nonsync_existence_stability}. When the fixed point loses stability through a transcritical bifurcation, the stable attractor becomes a fixed point of the form $(\phi^x,0,0)$, where $\phi^x\neq 0$. For $\mu^y \approx 1.4$, the stable solution approximately takes the form $(-T/2,0,0)$, indicating that the excitatory population is stable near anti-phase.
}

\begin{figure}
\centering
 \includegraphics[width=.75\textwidth]{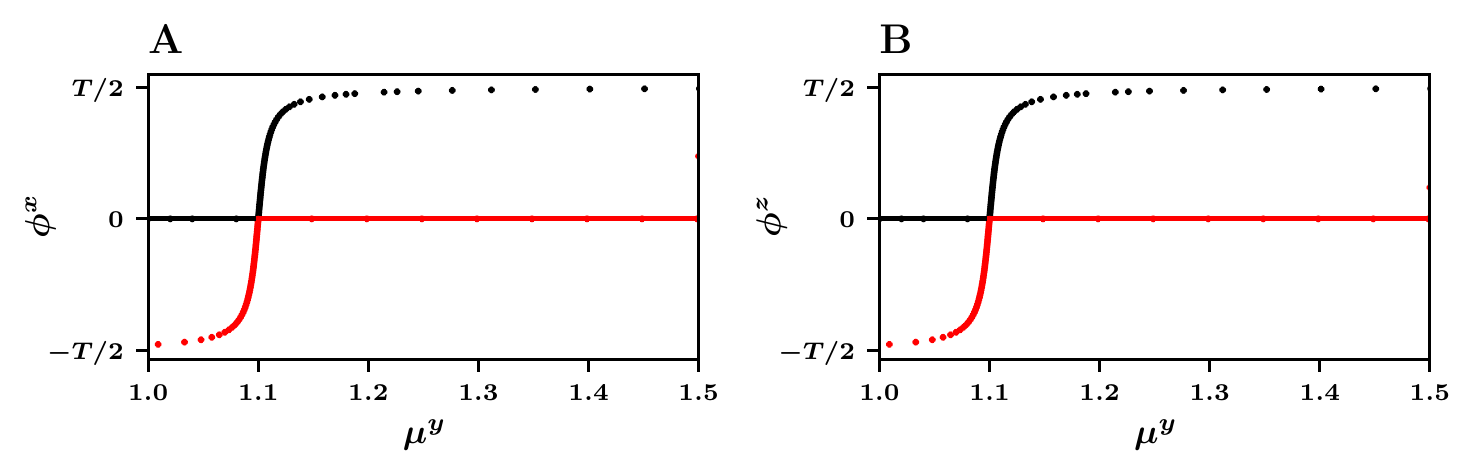}
 \caption{\editc{Stability analysis of phase-locked solutions of the form $(\phi^x,0,\phi^z)$ using parameters from Figures \ref{fig:micro_vs_macro} and \ref{fig:h_theta}. Black dots and lines: stable fixed points. Red dots and lines: unstable fixed points. A: x-coordinate values of fixed points. B: z-coordinate values of fixed points.}}\label{fig:two_nonsync_existence_stability}
\end{figure}

\editc{We now turn to the final stable branch, which takes the form $(\phi^x,0,\phi^z)$ (Figure \ref{fig:two_nonsync_existence_stability}). In panel A, we show the $\phi^x$ coordinate value as a function of $\mu^y$ and panel B shows the $\phi^z$ coordinate value as a function of $\mu^y$. Initially, synchrony is stable, until the bifurcation at $\mu^y = 1.1$, which leads to a stable branch that asymptotically approaches anti-phase as a function of $\mu^y$, and an unstable branch at the origin. We used \texttt{XPPAUT} to follow the equilibria as a function of $\mu^y$. There exist no other stable fixed points, concluding our analysis of existence and stability in the case of the fixed mean.}

\subsection{Existence and Stability of Phase-Locked Solutions (Slowly Varying Mean)}

\begin{figure}
\centering
 \includegraphics[width=.75\textwidth]{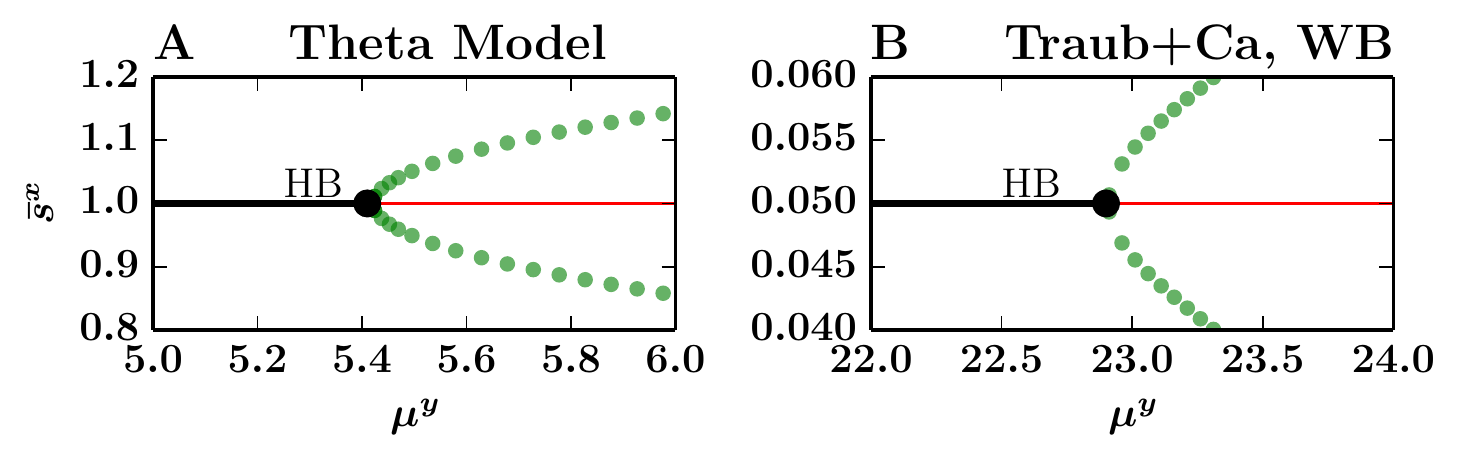}
 \caption{\edita{Hopf bifurcations in the mean field \yp{(Equation \eqref{eq:meanfield})}. A: Supercritical Hopf bifurcation in the mean field of the theta network Equations \eqref{eq:theta_avg1} and \eqref{eq:theta_avg2}. Parameters $a^x=0.5, b^x=7, c^x=6.5, a^y=1.1, b^y=25, c^y=25.1, \mu^x=1$. Black: stable fixed point. red solid: unstable fixed point. Green: stable periodic solution. B: Supercritical Hopf bifurcation in the mean field of the Traub+Ca and Wang-Buzs{\'a}ki network. Parameters \editb{$I^{xx} = 101.5\mu\text{A}/\text{cm}^2, I^{xy}=104\mu\text{A}/\text{cm}^2, I^{yx}=13\mu\text{A}/\text{cm}^2, I^{yy}=10.5\mu\text{A}/\text{cm}^2, \mu^x=1\text{ms}$}}}\label{fig:supercrit}
\end{figure}

\edita{With particular coupling parameter choices, the mean field undergoes a supercritical Hopf bifurcation and gives rise to slow, stable oscillations (Figure \ref{fig:supercrit}A). This slowly varying mean has the effect of forcing the excitatory population to spike at a different frequency from the inhibitory population. The goal of this section is to analyze the existence and stability of fixed points \editc{of the phase model} in this case. 

\subsubsection{Hopf Bifurcation in the Slowly Varying Case}
Figure \ref{fig:theta_full_vs_theory}A} shows slow, periodic behavior in the mean synaptic values. This periodic solution is a stable limit cycle solution arising from a supercritical Hopf bifurcation. By using the mean field in Equations \eqref{eq:theta_avg1} and \eqref{eq:theta_avg2}, we show existence of of a Hopf bifurcation and its criticality numerically.

\subsubsection{Phase Models Modulated by Slowly Varying Synapses}

\begin{figure}
\centering
 \includegraphics[width=.75\textwidth]{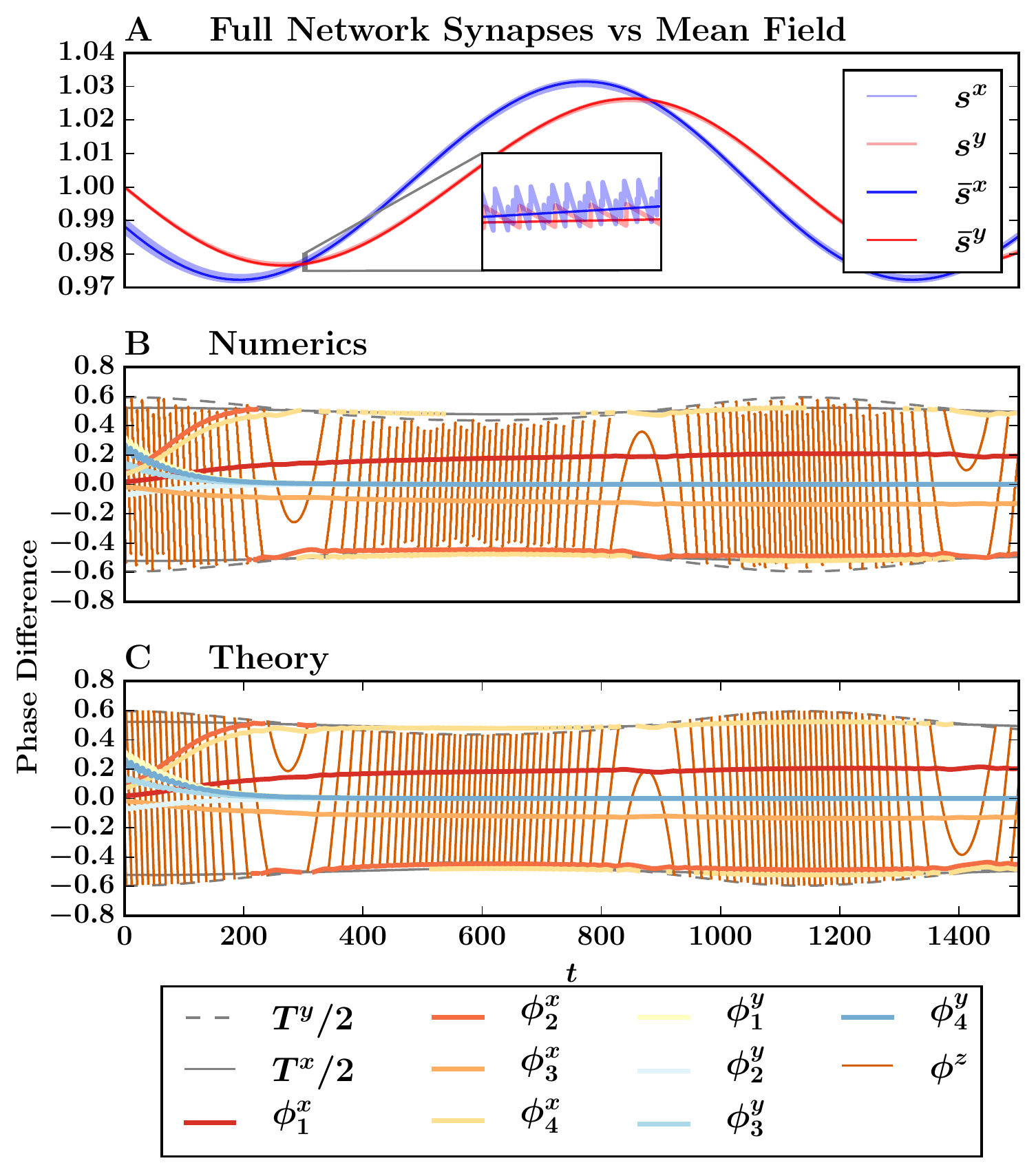}
 \caption{Numerics and theory for the theta model with slowly varying synapses \yp{for $N=5$}. \edita{A}: Mean field solutions $\bar s^{x,y}$ (solid lines) vs full network synaptic variables $s^{x,y}$ (opaque lines). Inset: Example of mean field solution plotted over the numerically simulated synaptic solutions. \edita{B}: Theta model simulation. \edita{C}: Phase model simulation. \yp{The legend at the bottom indicates which variables correspond to which colors}. Gray solid and gray dashed lines denote antiphase values $T^x/2$ and $T^y/2$ over time, respectively. Parameters as in Figure \ref{fig:supercrit} and $\ve = 0.005$.}\label{fig:theta_full_vs_theory}
\end{figure}

\edita{Figure \ref{fig:theta_full_vs_theory}B} shows the results of the numerical simulation in terms of phase differences for \yp{$N=5$ neurons per population}. Due to the slowly varying synaptic variables, the period of the oscillators change (as shown by the dashed gray and solid gray anti-phase lines). \yp{Generally, the phase differences in the excitatory population, $\phi^x_1,\ldots,\phi^x_4$, tend toward non-synchronous phase-locked activity}. \yp{In contrast, the phase differences in the inhibitory population, $\phi^y_1,\ldots,\phi^y_4$, tend toward synchrony. The difference in periods of the oscillators contributes to the phase drift between populations, quantified by $\phi^z$ (orange).}

\edita{Figure \ref{fig:theta_full_vs_theory}C} shows the results of the phase model simulation in terms of \yp{the same set of} phase differences. We see the same general trends. \yp{Excitatory neurons tend to non-synchronous phase-locked solutions, inhibitory neurons tend to synchronize, and there exists a large phase drift between the populations.}
%

\editb{
\subsection{Theta Models with Input Heterogeneities}
In this section, we consider the same theta neurons as above with $N=2$ with an additional input heterogeneity:
\begin{equation}\label{eq:theta_network_het}
\begin{split}
 \frac{dx_j}{dt} &= \pi(1-\cos(x_j) + (1+\cos(x_j))[(a^x+\ve \eta_j^x)+b^x s^x - c^x s^y]),\\
 \frac{dy_j}{dt} &= \pi(1-\cos(y_j) + (1+\cos(y_j))[(a^y+\ve \eta_j^y)+b^y s^x - c^y s^y]),\\
 \mu^x \frac{ds^x}{dt} &= \ve \left[-s^x + \frac{1}{N} \sum_j \delta(x_j - \pi)\right],\\
 \mu^y \frac{ds^y}{dt} &= \ve \left[-s^y + \frac{1}{N} \sum_j \delta(y_j - \pi)\right].
\end{split}
\end{equation}

We place no restriction on the heterogeneities $\eta^k_j$, so long as they are chosen such that $\ve \eta^k_j$ remains order $\ve$. In this example, we draw $\eta^k_j$ from a uniform distribution on the interval $[-1,1]$. With the numpy \cite{python} random seed set to 0, the four randomly chosen numbers are \texttt{[0.09762701, 0.43037873, 0.20552675, 0.08976637]}. We show an example of a simulation in Figure \ref{fig:theta_input_het}.

\begin{figure}
 \centering
 \includegraphics[width=.75\textwidth]{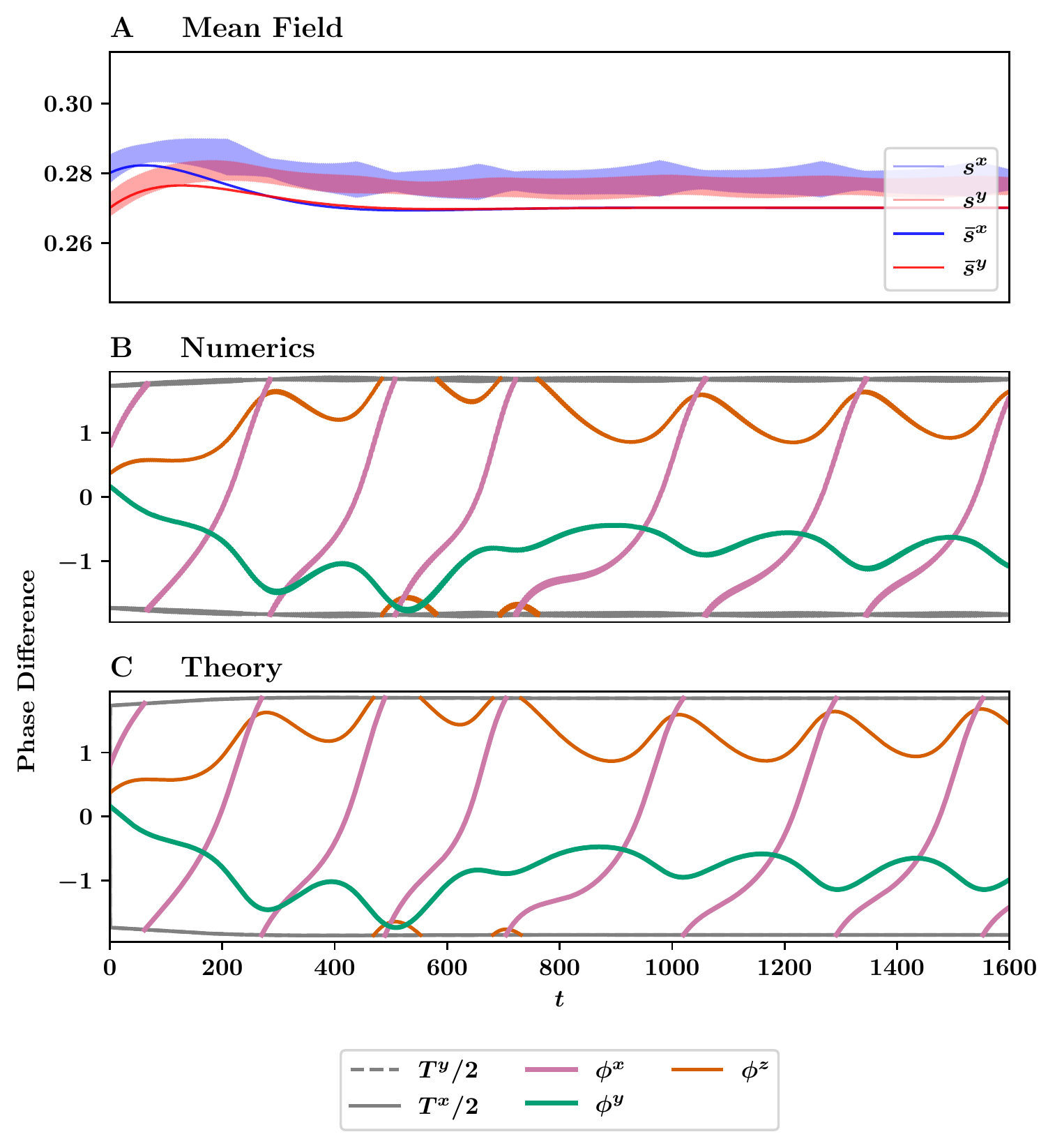}
 \caption{\editb{Effects of input heterogeneity. A: The full network simulation of the \yp{synaptic variables} (transparent blue, red labeled $s^{x,y}$) and the mean field approximation without input heterogeneities (solid blue, red labeled $\bar s^{x,y}$). B: Oscillator phase differences in the full network simulation. C: Predicted oscillator phase differences in the reduced model. Parameter values are the same as in Figure \ref{fig:h_theta} and Figure \ref{fig:micro_vs_macro}, with $\mu^y=1.5$, $\ve = 0.01$.}}\label{fig:theta_input_het}
\end{figure}

In Figure \ref{fig:theta_input_het}A, the full network simulation of the \yp{synaptic variables} (transparent blue, red labeled $s^{x,y}$) differs slightly in mean from the mean field approximation without input heterogeneities (\yp{time-averaged synaptic variables} solid blue, red labeled $\bar s^{x,y}$). In Figure \ref{fig:theta_input_het}B, the input heterogeneity results in a phase drift in the excitatory population (pink). Figure \ref{fig:theta_input_het}C shows our proposed theory, which accurately captures the transient dynamics as well as the phase drift in the excitatory population.

}

\subsection{Wang-Buzs{\'a}ki and Traub with Calcium}
We now repeat the analysis above using biophysically realistic models. In this section, we consider the synchronization properties in two populations of excitatory and inhibitory conductance-based models. The Excitatory population consists of the Traub model \cite{traub1982simulation} with calcium current, while the inhibitory population consists of the Wang Buzs{\'a}ki model \cite{wang1996gamma}. As in the previous section, we consider two cases. In the first case, the synaptic mean values are fixed, and in the second case, the synaptic mean values are slowly varying. 

The Traub model with calcium is defined by the system
\begin{equation}\label{eq:tb}
 \begin{split}
\dot \x = \frac{d}{dt}\left(\begin{matrix}
  V\\
  x\\
  w\\
  [\text{Ca}]
\end{matrix}\right)
=\left(\begin{matrix}
  (-I_{\text{ionic}}  + I_{\text{ext}})/C\\
  a_x(V)(1-x)  - b_x(V)x\\
  (w_\infty(V)-w)/\tau_w(V)\\
  (-\alpha I_{\text{Ca}}-[\text{Ca}]/\tau_{\text{Ca}})
 \end{matrix}\right) = \F^{x}(\x,I_{\text{ext}})
 \end{split}
\end{equation}
where $x$ represents the dynamics of gating variables $h,m$, and $n$. The ionic currents are listed in Equation \eqref{eq:tb_ionic} of Appendix \ref{a:tb}.

The Wang-Buzs{\'a}ki system is given by
\begin{equation}\label{eq:wb}
 \begin{split}
\dot \y = \frac{d}{dt}\left(\begin{matrix}
  V\\
  x\\
\end{matrix}\right)
=\left(\begin{matrix}
  -I_{\text{ionic}}  + I_{\text{ext}}\\
  \phi(x_\infty - x)/\tau_x\\
 \end{matrix}\right) = \F^{y}(\y,I_{\text{ext}}),
 \end{split}
\end{equation}
where $x$ represents the dynamics of gating variables $h$ and $n$. The ionic currents are listed in Equation \eqref{eq:wb_ionic} of Appendix \ref{a:tb}.

We introduce coupling through \editc{currents}:
\begin{equation}\label{eq:tbwb_coupling}
\begin{split}
 \frac{d\x_i}{dt} &= \F^{x}(\x_i,I^{x} + \editb{I^{xx} s^{x} - I^{xy} s^{y}}) ,\\
 \frac{d\y_i}{dt} &= \F^{y}(\y_i,I^{y} + \editb{I^{yx} s^{x} - I^{yy} s^{y}}),\\
 \mu^x\frac{ds^x}{dt} &= \ve\left[-s^x + \frac{1}{N}\sum_{i=1}^N \yp{\sum_j} \delta(t-t\yp{^x_{i,j}})\right],\\
 \mu^y\frac{ds^y}{dt} &= \ve\left[-s^y + \frac{1}{N}\sum_{i=1}^N \yp{\sum_j} \delta(t-t\yp{^y_{i,j}})\right],
\end{split}
\end{equation}
where just as in Equations \eqref{eq:main1a}--\eqref{eq:main2b}, \yp{$t^x_{i,j}$ ($t^y_{i,j}$) is the time of the $j^\text{th}$ spike of neuron $i$ in population $\x$ ($\y$)}.

\editc{\textbf{Aside}: While we also could include synaptic coupling using conductance-based synapses, the mean field equations are more complex as they are not just functions of sums of excitatory and inhibitory currents. Thus, we will use the simpler type of coupling shown in Equation \eqref{eq:tbwb_coupling}.}

The synapses $s^x,s^y$ \editb{(dimensionless)} increment each time the voltage variable of the neural models cross $V=0$ from negative to positive. Unless otherwise stated, we choose $\yp{s^*} = 1/T = 0.05$ cycles/ms = $50$ \editc{Hz}.

\begin{figure}
 \centering
 \includegraphics[width=.75\textwidth]{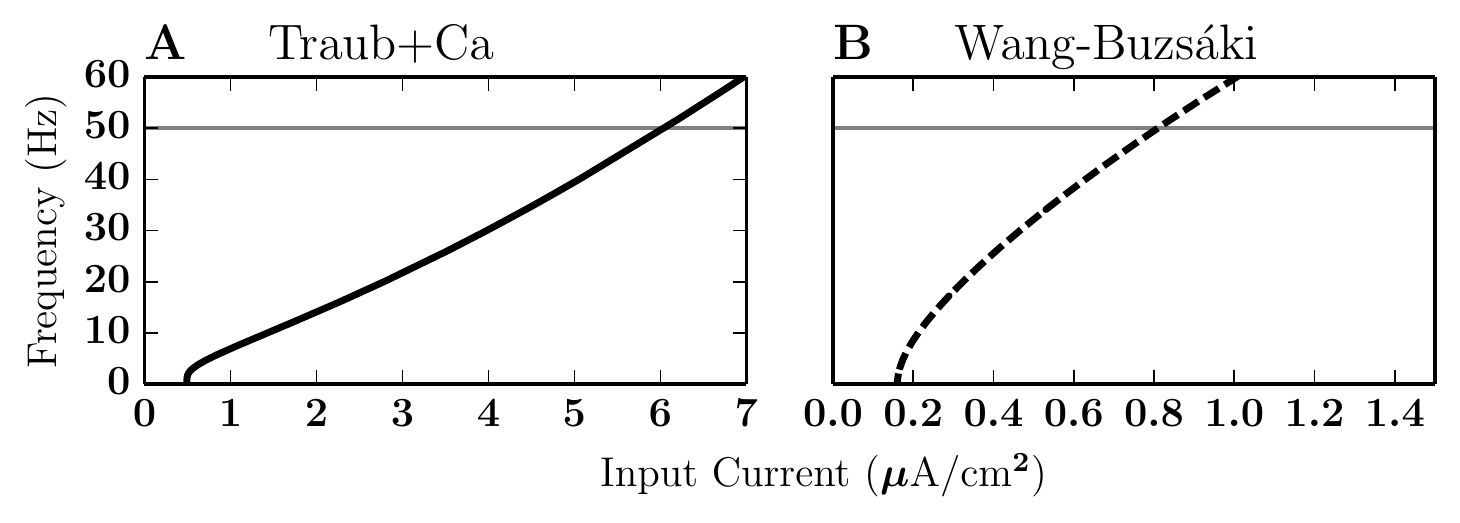}
 \caption{The frequency-current (FI) curves of the Traub with calcium model (solid black) and Wang-Buzs{\'a}ki model (dashed black). Frequency is in units of cycles per \yp{second}. The horizontal gray line through the frequency value $0.05$ denotes our choice of fixed mean synaptic current. Small and slow oscillations of the synaptic variables in this network are about this mean value and are fully determined by the values of these FI curves.}\label{fig:tbwb_fi}
\end{figure}

The mean field dynamics obey Equation \eqref{eq:meanfield}, where $\editb{\omega}^x$ is given by the frequency-input current (FI) function \edita{shown} by the black curve in Figure \ref{fig:tbwb_fi} and $\editb{\omega}^y$ is \edita{shown} by the FI curve given by the dashed curve in the same figure. We compute both curves numerically using \texttt{XPPAUTO} \cite{ermentrout2002simulating}.

This choice of coupling in Equation \eqref{eq:tbwb_coupling} results in scalar derivatives:
\begin{align*}
 \F^{x}_{s^{x}}(\Phi^x(t),\yp{s^*}, \yp{s^*}) &= (\editb{I^{xx}},0,0,0,0,0,0)^T,\\
 \F^{x}_{s^{y}}(\Phi^x(t),\yp{s^*}, \yp{s^*}) &= (-\editb{I^{xy}},0,0,0,0,0,0)^T,\\
 \F^{y}_{s^{x}}(\Phi^y(t),\yp{s^*}, \yp{s^*}) &= (\editb{I^{yx}},0,0)^T,\\
 \F^{y}_{s^{y}}(\Phi^y(t),\yp{s^*}, \yp{s^*}) &= (-\editb{I^{yy}},0,0)^T.
\end{align*}
Thus the $H$ functions of Equations \eqref{eq:theta_x} and \eqref{eq:theta_y} for this system are given by
\begin{equation}\label{eq:tbwb_h}
 \begin{split}
 H^{xx}(\phi) &= \frac{\editb{I^{xx}}}{T\mu^x} \int_0^{T} Z^x(t,\yp{s^*}) f(t+\phi) dt,\\
 H^{xy}(\phi) &= -\frac{\editb{I^{xy}}}{T\mu^y} \int_0^{T} Z^x(t,\yp{s^*}) f(t+\phi) dt,\\
 H^{yx}(\phi) &= \frac{\editb{I^{yx}}}{T\mu^x}\int_0^{T} Z^y(t,\yp{s^*}) f(t+\phi) dt,\\
 H^{yy}(\phi) &= -\frac{\editb{I^{yy}}}{T\mu^y}\int_0^{T} Z^y(t,\yp{s^*}) f(t+\phi) dt.
 \end{split}
\end{equation}

\begin{figure}
 \centering
 \includegraphics[width=.75\textwidth]{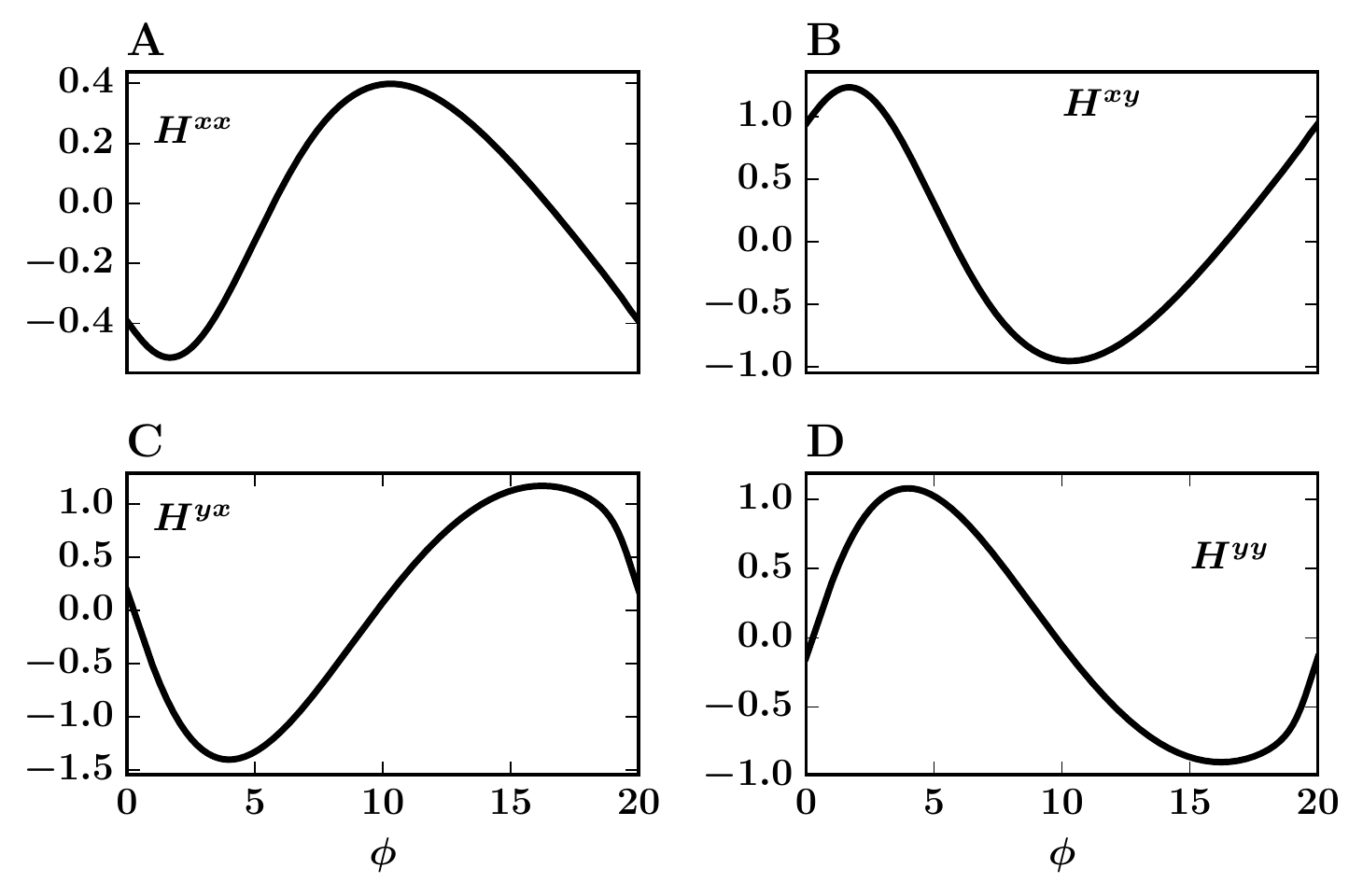}
 \caption{The $H$ functions of the Traub with calcium and Wang-Buzs{\'a}ki network. \edita{A: $H^{xx}$. B: $H^{xy}$. C: $H^{yx}$. D: $H^{yy}$. Parameter values \editb{$I^{xx} = 10\mu\text{A}/\text{cm}^2, I^{xy}=14\mu\text{A}/\text{cm}^2, I^{yx}=13\mu\text{A}/\text{cm}^2, I^{yy}=10\mu\text{A}/\text{cm}^2, \mu^x=\mu^y=1\text{ms}$}}.}\label{fig:h_fun_tbwb}
\end{figure}

We show plots of these $H$ functions in Figure \ref{fig:h_fun_tbwb}.

\begin{figure}[ht!]
\centering
 \includegraphics[width=.75\textwidth]{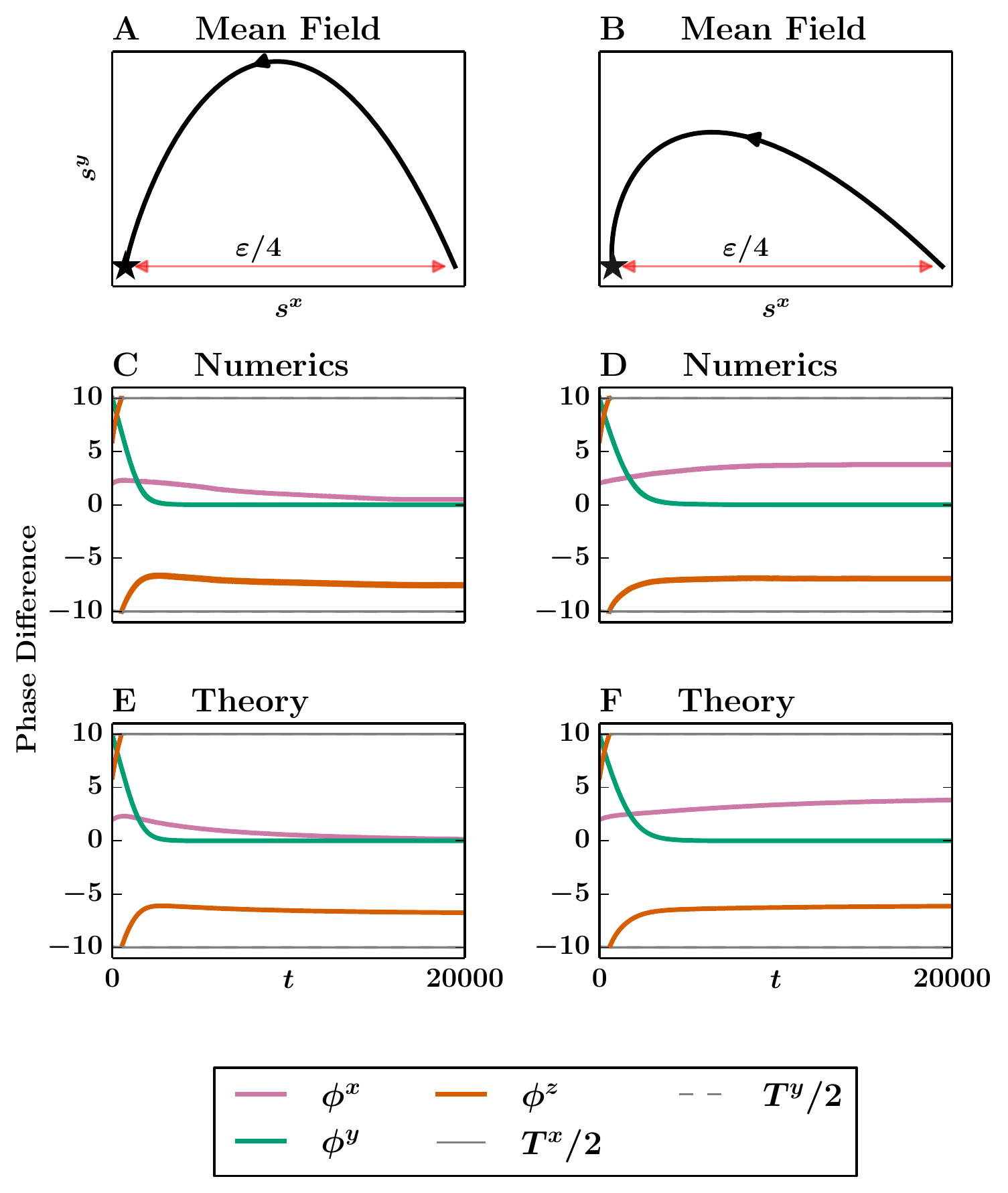}
 \caption{\edita{Mean field description and microscopic behavior with constant mean synapses. A: mean synaptic variables (\yp{$\bar s^{x,y}$}, black). The star denotes a stable fixed point. C: phase difference at the spiking level in the full conductance-based model simulation. The estimated half-period is shown in gray solid ($T^x/2$) and gray dashed ($T^y/2$) (they happen to overlap substantially and are almost impossible to distinguish). E: phase difference using our proposed phase reduction. Parameters are the same as in Figure \ref{fig:h_fun_tbwb}, and $\mu^x=\mu^y=1$. In the right column, we increase $\mu^y=1$ to $\mu^y = 2.5$ and plot the same data in the same order with the same initial conditions. Other parameters: $\ve = 0.0025$, \editb{$I^x=6.74\mu\text{A}/\text{cm}^2$, $I^y=0.66\mu\text{A}/\text{cm}^2$.}}}\label{fig:micro_vs_macro_tbwb}
\end{figure}

In Figure \ref{fig:micro_vs_macro_tbwb}, we simulate 2 excitatory Traub with calcium conductance-based models (Traub with calcium, Equation \eqref{eq:tb}), and 2 inhibitory conductance-based models (Wang-Buzs{\'a}ki, Equation \eqref{eq:wb}) with constant mean-field dynamics. All parameter values are the same except the parameter $\mu^y=1$ (left column) and $\mu^y=2.5$ (right column).

We plot the mean field in \edita{panels A and B} using the same scale to emphasize the qualitative difference in the mean field description. The stability remains the same between the left and right columns (negative real eigenvalues in both cases). The \edita{double-headed} red arrow indicates the magnitude of the perturbation off the fixed point. In both columns we choose to perturb the $s^x$ variable by magnitude $\ve/4$, where $\ve=0.0025$. Interestingly, this system exhibits \edita{similar features} in the mean field description shown in Figure \ref{fig:micro_vs_macro}, and the microscopic dynamics reach different steady-states despite no detectable changes to the stability of the mean field model.

\edita{Panel C shows that} the excitatory ($\phi^x$, purple) and inhibitory populations ($\phi^y$, green) approach synchrony. In \edita{panel D}, we re-initialize the simulation with the same initial conditions for all variables with only one change in the synaptic time constant from $\mu^y=1$ to $\mu^y=2.4$. The excitatory population reaches a non-synchronous steady-state phase locked value, \editb{indicating nearly a quarter-period difference in spike times}. \edita{Panels E and F show that our theory correctly predicts the differing steady state dynamics in panels C and D, respectively.}


\subsubsection{Existence and Stability of Phase-Locked Solutions (Fixed Mean)}
\begin{figure}[ht!]
\centering
 \includegraphics[width=.75\textwidth]{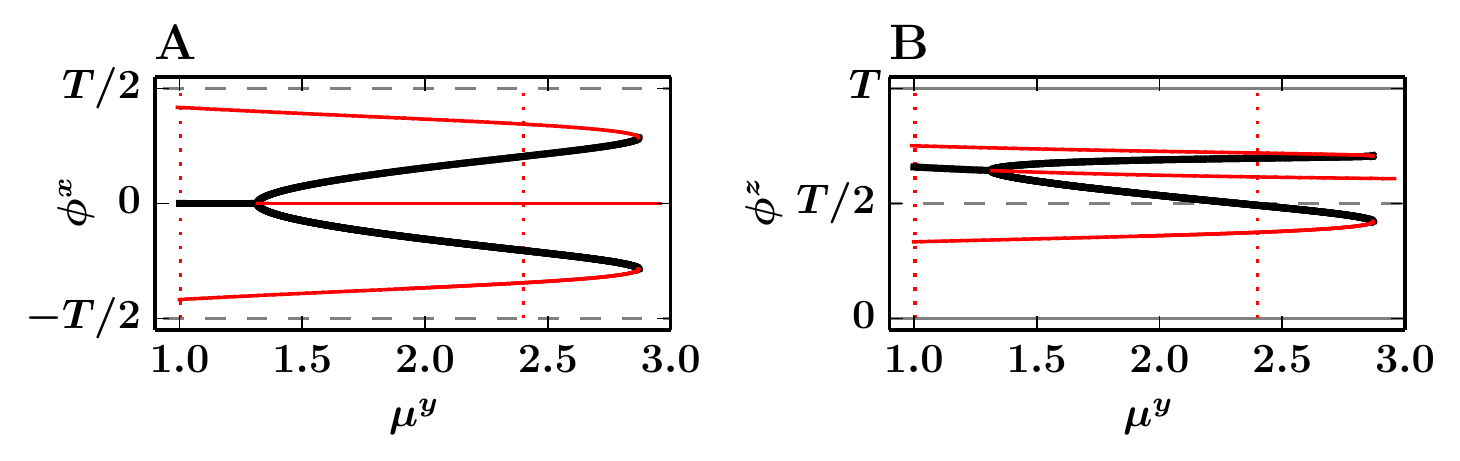}
 \caption{Existence and stability of phase locked solutions in the Traub Wang-Buzs{\'a}ki network. \edita{Vertical red dotted lines denote the two parameter values corresponding to the left and right columns of Figure \ref{fig:micro_vs_macro_tbwb}. Black curves denote stable values and solid red curves denote unstable values. A: Branches of $\phi^x$ fixed point values. B: Branches of $\phi^z$ fixed point values. Generally, $\phi^y=0$ (data not shown). Note that for $\mu^y=1$, the only fixed point that exists agrees with the steady-state in Figure \ref{fig:micro_vs_macro_tbwb}. For $\mu^y=2.4$, the fixed point corresponding to the lower stable branch of panel A and the upper stable branch of panel B coincides with the right column of Figure \ref{fig:micro_vs_macro_tbwb}}. Parameter values are identical to Figure \ref{fig:micro_vs_macro_tbwb}.}\label{fig:tbwb_existence_stbl}
\end{figure}

We now analyze the phase locked solutions of this system in the case of constant-mean synapses. As in the network of theta neurons, we use coupling parameters that lead to changes in the synchronization properties of the oscillators as a function of $\mu^y$, while the mean field remains invariant. We show the existence and stability of \edita{phase locked solutions} of Figure \ref{fig:micro_vs_macro_tbwb} in Figure \ref{fig:tbwb_existence_stbl}.

\edita{In Figure \ref{fig:tbwb_existence_stbl} we plot the value of each coordinate as a function of $\mu^y$ ($\phi^x$ in panel A and $\phi^z$ in panel B). We do not show $\phi^y$ because $\phi^y=0$ for this parameter range. As expected, the point $(\phi^x,\phi^y,\phi^z)\approx(0,0,3T/4)$ is stable for $\mu^y=1$. As we increase $\mu^y$, the system undergoes a pitchfork bifurcation, resulting in two stable fixed points. The fixed point we see in Figure \ref{fig:micro_vs_macro_tbwb} corresponds to the upper branch of both panels, where $(\phi^x,\phi^y,\phi^z)\approx(T/4,0,3T/4)$}.

\subsubsection{Phase Locked Solutions (Slowly Varying Mean)}
\begin{figure}[ht!]
\centering
 \includegraphics[width=.75\textwidth]{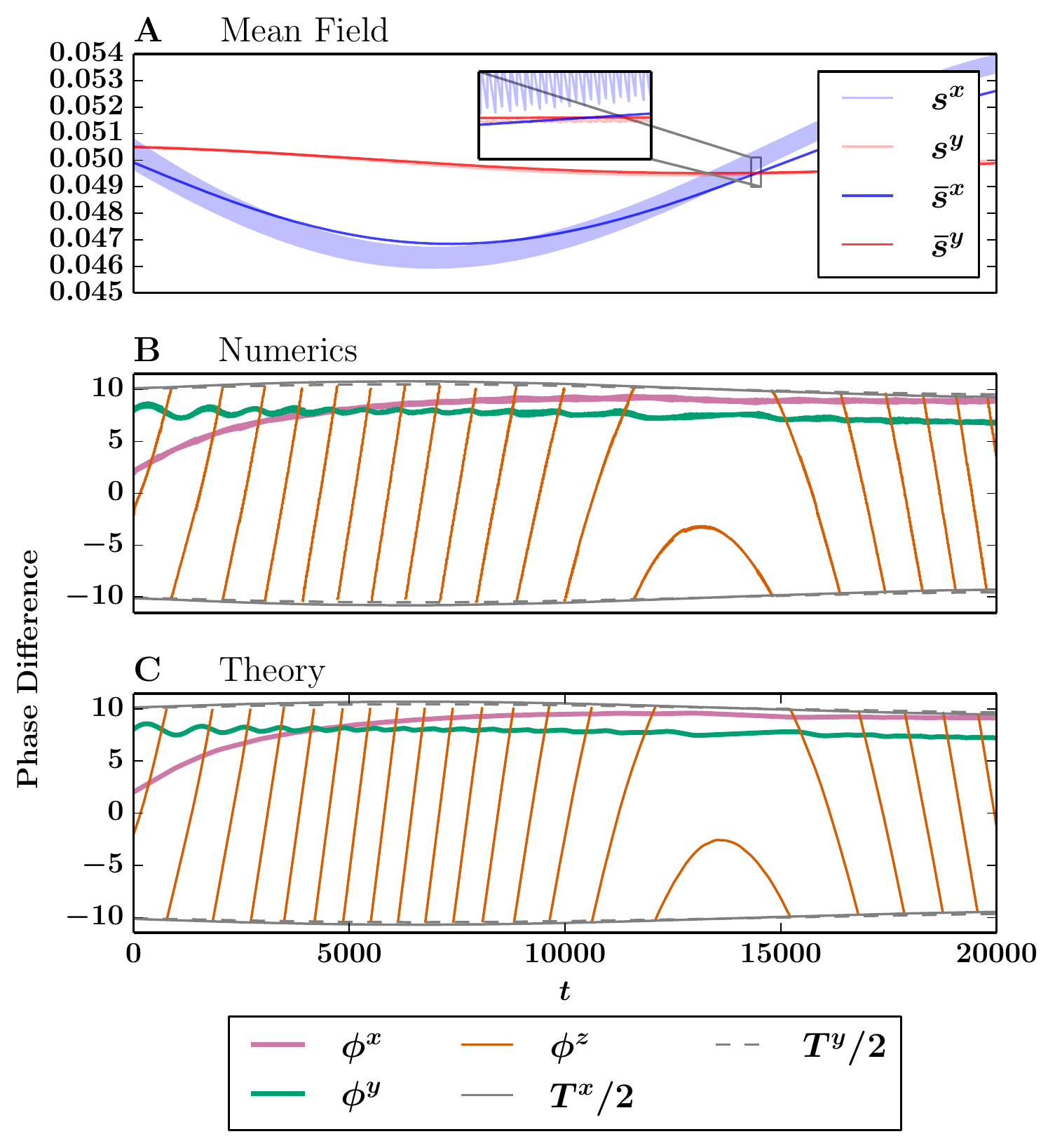}
 \caption{Numerics and theory for the Wang-Buzs{\'a}ki and Traub with calcium models with nonconstant mean synapses \yp{$\bar s^{x,y}$}. \edita{A}: Mean field solutions (\yp{$\bar s^{x,y}$}, solid lines) vs numerical synaptic variables (\yp{$s^{x,y}$}, opaque lines). Inset: Example of mean field solution plotted over the numerically simulated synaptic solutions. \edita{B}: Theta model simulation. \edita{C}: Phase model simulation. \edita{Parameters \editb{$I^{xx} = 101.5\mu\text{A}/\text{cm}^2, I^{xy}=104\mu\text{A}/\text{cm}^2, I^{yx}=13\mu\text{A}/\text{cm}^2, I^{yy}=10.5\mu\text{A}/\text{cm}^2, \mu^x=1\text{ms}, \mu^y=24.79\text{ms}$, \editd{$\ve = 0.00125$}.}}}\label{fig:tbwb_full_vs_theory}
\end{figure}

Finally, as in the theta network, the mean field of the Traub+Ca and Wang-Buzs{\'a}ki network may undergo a supercritical Hopf bifurcation \edita{(Figure \ref{fig:supercrit}B)}. In this section, we demonstrate that our theory accurately predicts the phase locking properties in this case of a slowly varying mean (Figure \ref{fig:tbwb_full_vs_theory}). We show the synaptic variables and mean field approximations in the top panel, the full numerical simulation in the middle panel, and our proposed theory in the bottom panel. We find that our theory correctly predicts the general trend of $\phi^x$ (pink) which tends towards antiphase, and of $\phi^y$ (green) which remains close to its initial condition.


\section{Discussion}
In the current study, we have shown that in an all-to-all, homogeneously coupled network of heterogeneous oscillators, there are two cases of the mean synaptic values that make the oscillators amenable to a phase reduction. In the first case, the mean synaptic values are fixed and equal, or fixed and different up to a small difference of order $\ve$. In the second case, the mean synaptic values are slowly varying with small amplitude up to order $\ve$. Using neurophysiologically motived models, we demonstrate that the phase reduction is accurate for at least order $1/\ve$ time. Moreover, we are able to explore the existence and stability of phase locked solutions in both cases.

Our choice of coupling in the Traub, Wang-Buzs{\'a}ki network also simplifies the analysis, but we \edita{need not} restrict the form of the input current. In fact, a biophysically realistic synaptic input current of the form $s^k g (V-V^k)$, where $g$ represents a conductance, $V^k$ the reversal potential, and $s^k$ the synaptic variable, is well within the scope of this study. In this case, we would add more complexity to the $H$ functions in Equation \eqref{eq:tbwb_h} \editc{and in the mean field equations}, but the analysis remains \editc{otherwise} unchanged.

\edita{Particular elements and motives of the current study are similar to existing works. \editb{Early studies in bridging spiking models to the mean field description use leaky integrate-and-fire (LIF) models with Poisson statistics \cite{amit1997model,amit1997dynamics}. Later studies derive additional statistics like the coefficient of variation in the interspike interval \cite{renart2007mean}. However, deriving equations measuring the degree of synchrony in a population of neurons (the order parameter \cite{Kuramoto1987}) is more recent. This reduction was first shown in \ypb{Ott and Antonsen 2008}\cite{ott2008low}, where the authors use what is now called the Ott-Antonsen ansatz to reduce \editc{an infinite number} of Kuramoto oscillators into a simple pair of differential equations for the order parameter.}

In \ypb{Montrio et al.~2015}\cite{montbrio_macroscopic_2015}, the authors derive a pair of ordinary differential equations for other macroscopic observables like the mean membrane potential and the firing rate by starting at the spiking level. They then show that the network is also amenable to the order parameter reduction using the Ott-Antonsen ansatz. Thus, it is possible to derive a complementary set of equations describing the mean field activity and the associated degree of synchronization. This derivation has also been applied to theta neurons in \ypb{Coombes and Byrne 2016}\cite{coombes2016next}, where the authors derive a complementary set of ordinary differential equations describing the population firing rate and the degree of synchrony.

The excitatory-inhibitory network structure has also been studied previously. In \ypb{Roulet and Mindlin 2016}\cite{roulet_average_2016}, they use the Ott-Antonsen ansatz to derive low dimensional differential equations for the order parameters of networks of excitatory and inhibitory Alder units:
\begin{align*}
 \dot \theta_i &= \omega_i - \cos\theta_i + I(\theta_j,\tilde \theta_j),\\
 \dot{\tilde \theta}_i &= \tilde \omega_i - \cos \tilde \theta_i + \tilde I (\theta_j,\tilde \theta_j),
\end{align*}
where the untilded variables refer to units in the excitatory population and the tilded variables refer to units in the inhibitory population. The coupling functions are defined as
\begin{align*}
 I(\theta_j,\tilde \theta_j) = \frac{k_E}{N} \sum_{j=1}^N (1-\cos\theta_j) - \frac{k_I}{\tilde N}\sum_{j=1}^{\tilde N} (1 - \cos\tilde \theta_j),
\end{align*}
where \yp{$k_E,k_I > 0$} denote the coupling strengths, $N$ and $\tilde N$ denote the number of neurons in each of the two populations.

Roulet and Mindlin then derive low dimensional differential equations for the order parameters of excitatory and inhibitory theta models:
\begin{align*}
 \dot \theta_i &= 1-\cos\theta_i + (1+\cos(\theta_i))[\eta_i + I(\theta_j,\tilde \theta_j)],\\
 \dot{\tilde\theta}_i &= 1 - \cos\tilde\theta_i + (1 + \cos\tilde\theta_i))[\tilde \eta_i + \tilde I(\theta_j,\tilde\theta_j)],
\end{align*}
with the same coupling functions as above. The equations are similar to the theta model we consider in this paper, except that the mean of the input current is slaved to the fixed parameters $\eta_i$ and thus the mean can not drift over time.

\yp{Aspects of our results have been considered in various studies, which we now summarize.} In \ypb{So et al.~2008}\cite{so2008synchronization}, the authors consider the effects of time-varying coupling on the synchronization properties of a network of Kuramoto oscillators. In particular, they show that with sufficiently fast binary switching of coupling strengths, the network exhibits behavior characteristic of a static network. In contrast, our results apply only to slow, continuous changes in input current. \yp{Next, although slow synapses exist in neural networks, synapses are not generally slow. Indeed, the problem of slow synapses has been addressed in \ypb{Laing 2014, 2015}\cite{laing2014derivation,laing_exact_2015}, where he derives a mean-field description of theta models while allowing for arbitrarily fast synaptic transmission. However, our results are a step in two important directions. One, we give our synapses their own dynamics, as is often the case in chemical synapses (as opposed to gap junction synapses), and two, the synaptic variables are additionally allowed to vary independently. Granted, they are only allowed to vary within a small neighborhood of a fixed point, but to the best of our knowledge, this independence is novel and generalizes existing studies \cite{rubinrubin}.}

Another aspect of the our results that have been studied in the past includes pulse coupled oscillators. Paz\'o and Montbri\'o\cite{pazo2014low}, use the Winfree model with a smooth pulse-like coupling of the form $P(x) = a_n(1+\cos(x))^n$. Combined with the Ott-Antonsen ansatz, they derive a pair of differential equations for the order parameter. In \ypb{Chandra et al.~2017}\cite{chandra_modeling_2017}, the authors consider a network of theta models with similar pulse-like coupling and derive the order parameter using the Ott-Antonsen ansatz. In addition, they relax the all-to-all coupling hypothesis and apply the Ott-Antonsen ansatz to a randomly generated network given an arbitrary degree distribution. General network structure satisfying classic weak coupling assumptions are studied in \ypb{Kori et al.~2009}\cite{kori2009collective}. In \ypb{Laing 2018}\cite{Laing2018}, the author considers all-to-all pulse-like coupling of theta neurons with and without synaptic delay and derives the order parameter using the Watanabe-Strogatz ansatz \cite{watanabe1994constants,watanabe1993integrability}.


Generalizing the synaptic weights is also a natural next step of the current study.} In this case, the system becomes
\begin{align*}
 \frac{d\x_i}{dt} &= \F^x\left(\x_i,\sum_{j=1}^N w_{ij}^{xx} s_j^x, \sum_{j=1}^N w_{ij}^{xy} s_j^y\right),\\
 \frac{d\y_i}{dt} &= \F^y\left(\y_i,\sum_{j=1}^N w_{ij}^{yx}s_j^x,\sum_{j=1}^N w_{ij}^{yy}s_j^y\right),\\
 \mu_i^x\frac{ds_i^x}{dt} &= \ve\left[-s_i^x + \yp{\sum_j \delta(t-t^{x}_{i,j})}\right],\\
 \mu_i^y\frac{ds_i^y}{dt} &= \ve\left[-s_i^y + \yp{\sum_j \delta(t-t^{y}_{i,j})}\right], \quad i=1\ldots,N.
\end{align*}
\yp{The time values $t^x_{i,j}$ and $t^y_{i,j}$ are as in Equations \eqref{eq:main1a}--\eqref{eq:main2b}.} To ensure that all neurons have the same firing rate when the synaptic variables are constant and identical, we require that
\begin{equation*}
 \sum_j w_{ij}^{jk} = \bar w^{jk}, \quad i=1,\ldots,N,\ \text{for} \ j,k=x,y,
\end{equation*}
where $\bar w^{jk}$ is a constant for each $j,k=x,y$. We are then free to choose coupling types of the form $W_{ij} = K(|i-j|\Delta x)$, where $K$ is a typical even kernel, like a Gaussian or difference of Gaussians, and $\Delta x = 1/N$. This type of modification brings us closer to classic spatially distributed neural field models and the resulting system remains amenable to the methods of the current study. This direction also ties in with \ypb{Laing 2014, 2015}\cite{laing2014derivation,laing_exact_2015}, where bump-type solutions are shown to exist \textit{a priori} in large networks of theta models. Our method could show the same results with more general models.

The Ott-Antonsen ansatz is an undoubtedly powerful tool for understanding oscillator models. However, it has some limitations which the current \editb{paper} addresses directly. Our \editb{proposed} theory offers a general dimension reduction of a finite number of $N$- and $M$-dimensional coupled oscillators. While our theory tolerates only small heterogeneities, we place no restrictions on how the heterogeneities are distributed. \editd{However, our theory is restrictive} in that the interactions must be on a slow timescale.

Another natural next step to consider involves the effects of noise on synchronization. In \ypb{Nagai and Kori 2010}\cite{nagai2010noise}, where a network of these oscillators are driven by a common Gaussian noise signal, \editd{the authors analytically show noise-induced synchronization, and suggest that weak common noise generally promotes synchronization of weakly coupled oscillators. We have shown to a limited extent the effects of noise by introducing input heterogeneities drawn from a uniform distribution. However, analyzing the effects of a single time-dependent noisy input signal in our framework requires different techniques beyond the scope of this paper, and warrants close study in its own right.}

\newpage
\section*{Acknowledgments}
GBE and YMP were partially supported by NSF DMS 1712922, and the Andrew Mellon predoctoral fellowship.

\appendix

\section{Model Equations and Parameters}
\subsection{Traub With Calcium}\label{a:tb}

\begin{align*}
\dot \x = \edita{\frac{d}{dt}\left(\begin{matrix}
  V\\
  x\\
  w\\
  [\text{Ca}]
\end{matrix}\right)}
=\left(\begin{matrix}
  (-I_{\text{ionic}}  + I_{\text{ext}})/C\\
  a_x(V)(1-x)  - b_x(V)x\\
  (w_\infty(V)-w)/\tau_w(V)\\
  (-\alpha I_{\text{Ca}}-[\text{Ca}]/\tau_{\text{Ca}})
 \end{matrix}\right) = \F^{x}(\x,I_{\text{ext}})
\end{align*}
where $x$ represents the dynamics of gating variables $h,m$, and $n$, and 
\begin{equation}\label{eq:tb_ionic}
 \begin{split}
 I_{\text{ionic}} &= I_{\text{Na}} + I_\text{K} + I_{\text{Ca}} + I_{\text{ahp}} + I_\text{M}+g_\text{L} (V-E_\text{L})\\
 I_{\text{Na}} &= g_{\text{Na}}m^3h(V-E_{\text{Na}})\\
 I_{\text{K}} &= g_{\text{K}}n^4(V-E_\text{K})\\
 I_{\text{Ca}} &= g_{\text{Ca}} M_{L\infty}(V) (V-E_{\text{Ca}})\\
 I_\text{ahp} &= \frac{g_{\text{ahp}}[\text{Ca}](V-E_\text{K})}{[\text{Ca}]+K_\text{d}}\\
 I_\text{M} &= g_\text{M} w (V-E_\text{K})
 \end{split}
\end{equation}

\edita{The voltage variable $V$ has dimensions of mV, all currents are in dimensions of \editb{$\mu$A/cm$^2$}, time is in units of milliseconds, the variables $n,m,h$, and $w$ are dimensionless, and the variable [Ca] represents the intracellular calcium concentration in millimolar units. We show dimensions of all model parameters in Table \ref{tab:tb}.}

\begin{align*}
a_m(V) &=\frac{0.32(V+54)}{1-\exp(-(V+54)/4)}\\
b_m(V) &=\frac{0.28(V+27)}{\exp((V+27)/5)-1}\\
a_h(V) &=0.128 \exp(-(V+50)/18)\\
b_h(V) &=\frac{4}{1 + \exp(-(V+27)/5)}\\
a_n(V) &=\frac{0.032 (V+52)}{1 - \exp(-(V+52)/5)}\\
b_n(V) &=0.5 \exp(-(V + 50)/40)\\
\tau_w(V) &= \frac{\tau_w}{3.3\exp((V-V_{wt})/20) + \exp(-(V-V_{wt})/20)}\\
w_\infty(V) &= \frac{1}{1 + \exp(-(V-V_{wt})/10)}\\
\edita{M_{L\infty}(V)} &= \edita{1/(1+\exp(-(V-V_{\text{Lth}})/V_{\text{shp}}))}
\end{align*}

\begin{table}
\caption {Traub with calcium parameter values} \label{tab:tb}
\begin{center}
\begin{tabular}{l|l}
Parameter & Value\\
\hline
$C$ & 1 $\mu \text{F}/\text{cm}^2$\\
$E_K$& $-100\text{mV}$\\
$E_{Na}$& $50 \text{mV}$\\
$E_L$& $-67 \text{mV}$\\
$E_{\text{Ca}}$&$120\text{mV}$\\

$g_L$& $0.2 \text{mS}/\text{cm}^2$\\
$g_K$& $80 \text{mS}/\text{cm}^2$\\
$g_{Na}$& $100 \text{mS}/\text{cm}^2$\\
$g_{m}$& $0 \text{mS}/\text{cm}^2$\\

$g_{\text{Ca}}$& $1 \text{mS}/\text{cm}^2$\\
$g_{ahp}$& $0.5 \text{mS}/\text{cm}^2$\\

$K_d$ & $1 \text{mM}$\\
$\alpha$& $0.002\text{mmol}/(\text{cm} \times \text{nC})$\\
$\tau_{\text{Ca}}$& $80 \text{ms}$\\

$V_{\text{shp}}$&$2.5 \text{mV}$\\
$V_{\text{Lth}}$&$-25 \text{mV}$\\
$V_{\text{sshp}}$&$2 \text{mV}$\\
$V_{\text{th}}$&$-10 \text{mV}$\\

$V_{wt}$&$-35 \text{mV}$\\
$\tau_w$&$100 \text{ms}$\\
\end{tabular}
\end{center}
\end{table}

\subsection{Wang-Buzs{\'a}ki}\label{a:wb}
\begin{align*}
\dot \y = \edita{\frac{d}{dt}\left(\begin{matrix}
  V\\
  x\\
\end{matrix}\right)}
=\left(\begin{matrix}
  -I_{ionic}  + I_{\text{ext}}\\
  \phi(x_\infty - x)/\tau_x\\
 \end{matrix}\right) = \F^{y}(\y,I_{\text{ext}}),
\end{align*}
where $x$ represents the dynamics of gating variables $h$ and $n$, and 
\begin{equation}\label{eq:wb_ionic}
 \begin{split}
 I_\text{ionic} &= g_\text{L} (V-E_\text{L}) + I_\text{Na} + I_\text{K}\\
 I_\text{Na} &= g_{\text{Na}} m_\infty^3 h(V-E_{\text{Na}})\\
 I_\text{K} &= g_\text{K}n^4 (V-E_\text{K})
 \end{split}
\end{equation}
\edita{As in the Traub model above, the variable $V$ has dimensions of \editb{mV}, time units of milliseconds, the variables $h$ and $n$ are dimensionless, and currents are in units of \editb{$\mu$A/cm$^2$}. We show dimensions of all model parameters in Table \ref{tab:wb}.}

\begin{align*}
\alpha_m(V) &= \frac{0.1(V+35)}{1-\exp(-(V+35)/10}\\
\beta_m(V)  &= 4\exp(-(V+60)/18)\\
\alpha_h(V) &= 0.07\exp(-(V+58)/20)\\
\beta_h(V)  &= \frac{1}{1+\exp(-(V+28)/10)}\\
\alpha_n(V) &= \frac{0.01(V+34)}{1-\exp(-(V+34)/10)}\\
\beta_n(V)  &= 0.125\exp(-(V+44)/80)\\
x_\infty &= x_1/(x_1+x_2)\\
\tau_x &= 1/(x_1+x_2)
\end{align*}
where $x$ in the last two lines represents $m, h$, or $n$ and $x_1, x_2$ may be $\alpha_x$ and $\beta_x$, respectively.
\begin{table}
\caption {Wang-Buzs{\'a}ki parameter values} \label{tab:wb}
\begin{center}
\begin{tabular}{l|l}
Parameter & Value\\
\hline
$E_K$& $-90\text{mV}$\\
$E_{Na}$& $55 \text{mV}$\\
$E_L$& $-65 \text{mV}$\\

$g_L$& $0.1 \text{mS}/\text{cm}^2$\\
$g_{Na}$& $35 \text{mS}/\text{cm}^2$\\
$g_K$& $9 \text{mS}/\text{cm}^2$\\

$\phi$& $5$
\end{tabular}
\end{center}
\end{table}


\section{Derivation of Spiking Term}\label{a:spiking}
Recall our starting ansatz for the phase equation,
\begin{equation*}
\begin{split}
  \x_i(t,\tau) &= \x_i(t + \theta_i(\tau),\yp{s^*}) = \Phi^x(t + \theta_i^x(\tau),\yp{s^*}) + \varepsilon \xi_i^x(t + \theta_i^x(\tau),\yp{s^*}) + O(\varepsilon^2),\\
  \y_i(t,\tau) &= \y_i(t + \theta_i(\tau),\yp{s^*}) = \Phi^y(t + \theta_i^y(\tau),\yp{s^*}) + \varepsilon \xi_i^y(t + \theta_i^y(\tau),\yp{s^*}) + O(\varepsilon^2),\\
 s^x(t,\tau) &= \yp{s^*}(\tau) + \frac{\varepsilon}{N\mu^x} \sum_j f\left(t+\theta^x_j(\tau)\right) + O(\ve^2),\\
 s^y(t,\tau) &= \yp{s^*}(\tau) + \frac{\varepsilon}{N\mu^y} \sum_j f\left(t+\theta^y_j(\tau)\right) + O(\ve^2),
\end{split}
\end{equation*}
where \yp{$f$ represents the small-magnitude, fast-timescale effects of the variables $\x_i(t,\tau)$ and $\y_i(t,\tau)$ on the synaptic variables. In this section, we derive the order $\ve$ term $f$:}
\begin{align*}
 f(t+\theta) &= \left [ \left(1 - (t+\theta)/T \yp{\pmod{1}}\right) -1/2\right].
\end{align*}

For simplicity, consider a network consisting of one excitatory neuron $\x_1$ with one synaptic variable $s^x(t,\tau)$. Note that following a spike, the solution $\x_1$ increments by $\varepsilon_k \equiv \varepsilon/\mu^k$ and decays exponentially. Moreover, each $s^k$ is periodic with $s^k(T^+) = s^k(0)$, where $k=x,y$. Putting these facts together, we have that
\begin{equation*}
 s^k(T^+) = s^k(0)e^{-\varepsilon_k T} + \varepsilon_k = s^k(0).
\end{equation*}
Solving for $s^k(0)$ reveals
\begin{equation*}
 s^k(0) = \frac{\varepsilon_k}{1-e^{-\varepsilon_k T}}.
\end{equation*}
Therefore, $s^k(t)$ after a spike is
\begin{equation*}
 s^k(t) = \frac{\varepsilon_k}{1-e^{-\varepsilon_k T}}e^{-\varepsilon_k t}.
\end{equation*}
\yp{Using Taylor expansions, we can rearrange the equation as}
\begin{equation*}
 s^k(t) = \frac{\varepsilon_k}{1-e^{-\varepsilon_k T}}e^{-\varepsilon_k t} = \frac{1}{T} + \varepsilon_k f(t),
\end{equation*}
which after a trivial rearrangement yields
\begin{equation*}
 \varepsilon_k  f(t) = \frac{\varepsilon_k}{1-e^{-\varepsilon_k T}}e^{-\varepsilon_k t} - \frac{1}{T}.
\end{equation*}
Since $\varepsilon_k$ is small, we take a Taylor expansion of the exponential and simplify in a series of algebraic steps:
\begin{align*}
\varepsilon_k f(t) &= \frac{\varepsilon_k(1 - \varepsilon_k t + O(\varepsilon_k^2))}{1 - (1 - \varepsilon_k T + (\varepsilon_kT)^2/2 + O(\varepsilon_k^3)} - \frac{1}{T}\\
&= \frac{1 - \varepsilon_k t + O(\varepsilon_k^2)}{ T - \varepsilon_kT ^2/2 + O(\varepsilon_k^2)} - \frac{1}{T}\\
&= \frac{1}{T}\frac{1 - \varepsilon_k t + O(\varepsilon_k^2)}{ 1 - \varepsilon_k T/2 + O(\varepsilon_k^2)} - \frac{1}{T}\\
&= \frac{1}{T} \left(\frac{1 - \varepsilon_k t + O(\varepsilon_k^2)}{ 1 - \varepsilon_k T/2 + O(\varepsilon_k^2)} - \frac{1-\varepsilon_kT/2+O(\varepsilon_k^2)}{1-\varepsilon_kT/2+O(\varepsilon_k^2)}\right)\\
&=\frac{1}{T} \frac{\varepsilon_k T/2 - \varepsilon_k t + O(\varepsilon_k^2)}{ 1 - \varepsilon_k T/2 + O(\varepsilon_k^2)}\frac{1 + \varepsilon_kT/2 + O(\varepsilon_k^2)}{1 + \varepsilon_kT/2 + O(\varepsilon_k^2)}\\
&=\frac{1}{T} \frac{\varepsilon_k T/2 - \varepsilon_k t + O(\varepsilon_k^2)}{ 1 + O(\varepsilon_k^2)}\\
&\approx \frac{1}{T}\varepsilon_k(T/2 - t).
\end{align*}
Thus,
\begin{equation*}
 f(t) = \left(\frac{1}{2} - \frac{t}{T} \right),
\end{equation*}
over one period. For multiple periods, the resulting function is a sawtooth. In our implementations we write
\begin{equation*}
 f(t) = \left( (1 - t/T) \yp{\pmod{1}} \right) - 1/2,
\end{equation*}
\yp{because it is the most natural formulation for computer simulations.} In general, we need to account for possible slow timescale phase shifts $\theta^k_j(\tau)$ and the contributions from multiple spikes. We simply sum these contributions to arrive at the desired form:
\begin{equation*}
 \frac{\varepsilon}{N\mu^x} \sum_j f\left(t+\theta^x_j(\tau)\right).
\end{equation*}

\section{Fourier Coefficients}

\begin{table}[H]
\centering
\caption {\edita{H-function coefficients of the theta model. The series takes the form $\sum_{i=1}^n a_i \cos(i x) + b_i \sin(i x)$. Error = 7e-3.}} \label{tab:fourier_theta}
\begin{tabular}{l|l|l|l|l}
Coefficient &$H^{xx}$ & $H^{xy}$ & $H^{yx}$& $H^{yy}$  \\
\hline
$a_1$ & 0.006693442 & -0.00736278 & 0.006693442 & -0.00736278\\
$b_1$ & -1.09191412 & 1.201105540 & -1.09191412 & 1.201105540 
\end{tabular}
\end{table}

\begin{table}[H]
\centering
\caption {\edita{$H$-function coefficients of the Traub with calcium ($H^{xx}, H^{xy}$) and Wang-Buzs{\'a}ki ($H^{yx}, H^{yy}$). $a_0 + \sum_{i=1}^n a_i \cos(i x) + b_i \sin(i x)$. Maximum pointwise error = 1e-2.}} \label{tab:fourier_tbwb}
\begin{tabular}{l|l|l|l|l}
Coefficient &$H^{xx}$ & $H^{xy}$ & $H^{yx}$& $H^{yy}$  \\
\hline
$a_0$ & -0.00014179 & 0.000340298 &  & \\
$a_1$ & -0.40559133 & 0.973419204 & 0.014799680 & -0.01138436\\
$a_2$ & -0.00296584 & 0.007118019 & 0.120314654 & -0.09254973\\
$a_3$ & 0.008194908 & -0.01966778 & 0.045877649 & -0.03529049\\
$a_4$ & 0.003950934 & -0.00948224 & 0.016650161 & -0.01280781\\
$a_5$ & 0.001870675 & -0.00448962 & 0.004690780 & -0.00360829\\
$a_6$ & 0.000941694 & -0.00226006 & -0.00025889 & 0.000199148\\
$a_7$ & 0.000497815 & -0.00119475 & -0.00214311 & 0.001648549\\
$a_8$ & 0.000271530 & -0.00065167 & -0.00264425 & 0.002034040\\
$a_9$ & 0.000151095 & -0.00036262 & -0.00253634 & 0.001951034\\
$a_{10}$ & 8.560764307 & -0.00020545 & -0.00218981 & 0.001684474\\
\hline
$b_1$ & -0.14512093 & 0.348290252 & -1.29140450 & 0.993388080\\
$b_2$ & -0.07327966 & 0.175871190 & -0.22112111 & 0.170093167\\
$b_3$ & -0.01561869 & 0.037484866 & -0.07939645 & 0.061074193\\
$b_4$ & -0.00449755 & 0.010794122 & -0.03842115 & 0.029554737\\
$b_5$ & -0.00168564 & 0.004045538 & -0.02107670 & 0.016212846\\
$b_6$ & -0.00073921 & 0.001774105 & -0.01208433 & 0.009295643\\
$b_7$ & -0.00034753 & 0.000834072 & -0.00691601 & 0.005320008\\
$b_8$ & -0.00016011 & 0.000384282 & -0.00380417 & 0.002926287\\
$b_9$ & -6.11449933 & 0.000146747 & -0.00190524 & 0.001465571\\
$b_{10}$ & -5.59280546 & 1.342273311 & -0.00075672 & 0.000582099
\end{tabular}
\end{table}

\end{document}